\documentclass[twocolumn,appendixfloats,tighten]{aastex6}
\usepackage{newtxtext,newtxmath}
\usepackage[T1]{fontenc}
\usepackage{ae,aecompl}

\usepackage{amsmath}	
\usepackage{amssymb}	

\usepackage{ctable}
\usepackage{url}
\usepackage{xspace}
\usepackage[normalem]{ulem}
\usepackage{hyperref}
\usepackage[all]{hypcap}
\usepackage{graphicx}
\usepackage{comment}

\usepackage{amssymb}
\usepackage{pifont}

\usepackage{etoolbox}
\newtoggle{ApJFigs}
\togglefalse{ApJFigs}

\def\msun{{\rm\,M_\odot}}

\def\msun{{\rm\,M_\odot}}

\newcommand{\kms}{\, {\rm km\, s}^{-1}}

\newcommand{\be}{\begin{equation}}
\newcommand{\ee}{\end{equation}}

\newcommand{\au}{\mathrm{au}}

\def\h2{${\rm\,H_2}$}

\newcommand{\F}{Fig.}
\newcommand{\Fs}{Figs.}
\newcommand{\eq}{equation}

\newcommand{\ve}[1]{\boldsymbol{#1}}
\newcommand{\unit}[1]{\hat{\boldsymbol{#1}}}
\newcommand{\mci}{m_{i,\mathrm{c}}}
\newcommand{\cmark}{\ding{51}}%
\newcommand{\xmark}{\ding{55}}%

\newcommand{\fdm}{f_\mathrm{dm}}
\newcommand{\chie}{\chi_\mathrm{eff}}
\newcommand{\chip}{\chi_\mathrm{p}}

\newcommand{\hz}{\mathrm{Hz}}
\newcommand{\myr}{\mathrm{Myr}}
\newcommand{\yr}{\mathrm{yr}}
\newcommand{\gyr}{\mathrm{Gyr}}
\newcommand{\gpc}{\mathrm{Gpc}}
\newcommand{\pgpy}{\gpc^{-3}\,\yr^{-1}}

\usepackage{color}

\begin{document}

\title{Was GW190412 born from a hierarchical 3+1 quadruple configuration?}

\author{Adrian S. Hamers\altaffilmark{1} and Mohammadtaher Safarzadeh\altaffilmark{2}}
\altaffiltext{1}{Max-Planck-Institut f\"{u}r Astrophysik, Karl-Schwarzschild-Str. 1, 85741 Garching, Germany}
\altaffiltext{2}{Center for Astrophysics | Harvard \& Smithsonian, 60 Garden Street, Cambridge, MA, USA}

\begin{abstract} 
The gravitational wave source GW190412 is a binary black hole (BBH) merger with three unique properties: i) its mass ratio is about 0.28, the lowest found so far, ii) it has a relatively high positive effective spin parameter $\chie=0.25$, and iii) it is observed to be precessing due to in-plane projected spin of the binary with an in-plane precession parameter $\chip=0.3$. The two main formation channels of BBH formation fail to account for GW190412: field formation scenarios cannot explain the observed precession unless by invoking large natal kicks, and dynamical assembly in dense stellar systems is inefficient in producing such low mass-ratio BBH mergers. Here, we investigate whether `double mergers' in wide hierarchical quadruple systems in the `3+1' configuration could explain the unique properties of GW190412. In this scenario, a compact object quadruple system experiences two mergers: first, two compact objects in the innermost orbit merge due to secular chaotic evolution. At a later time, the merged compact object coalesces with another compact object due to secular Lidov-Kozai oscillations. We find that our scenario is consistent with GW190412. In particular, we find a preferential projected spin around $\chip=0.2$. However, the likelihood of a double merger is small and the formation efficiency of these systems is uncertain. If GW190412 originated from a double merger in a 3+1 quadruple, we find a strong constraint that the first merger likely occurred between roughly equal-mass BHs in the innermost orbit, since the recoil velocity from unequal-mass BHs would otherwise have disrupted the system. 
\end{abstract}

\section{Introduction}
\label{sec:intro}
Binary black hole (BBH) formation can broadly be divided into two categories: i) field formation in which two massive main sequence stars in a binary configuration end their lives leaving behind two BHs that will eventually merge by losing angular momentum due to the emission of the gravitational waves (GWs) (e.g., \citealt{1973NInfo..27...70T,1993MNRAS.260..675T,2002ApJ...572..407B,2003MNRAS.342.1169V,2007PhR...442...75K,
2012ApJ...759...52D,2012ApJ...757...27A,2013ApJ...779...72D,2014ApJ...789..120B,2016Natur.534..512B,2018MNRAS.473.4174Z,2018PhRvD..98h4036G,2019ApJ...870L..18Q,2020A&A...635A..97B,2020A&A...636A.104B,2020arXiv200411866O}), and ii) dynamical assembly in dense stellar systems such as globular clusters, open clusters, and nuclear star clusters, or in hierarchical systems (e.g., \citealt{1993Natur.364..423S,2000ApJ...528L..17P,2004Natur.428..724P,2006ApJ...637..937O,2012ApJ...757...27A,2011ApJ...741...82T,2014ApJ...781...45A,2014MNRAS.441.3703Z,2015ApJ...799..118P,2015PhRvL.115e1101R,2016MNRAS.459.3432M,2016MNRAS.460.3494S,2016MNRAS.463.2443K,2016ApJ...831..187A,2016ApJ...832L...2R,2016PhRvD..93h4029R,2017ApJ...836...39S,2017ApJ...836L..26C,2017ApJ...840L..14S,2017ApJ...841...77A,2017ApJ...846..146P,2018MNRAS.480L..58A,2018PhRvD..98l3005R,2018ApJ...855..124S,2018ApJ...853..140S,2018ApJ...865....2H,2018ApJ...856..140H,2018PhRvD..97j3014S,2018MNRAS.477.4423A,2018PhRvL.120o1101R,2018ApJ...853...93R,2018ApJ...856..140H,2018ApJ...860....5G,2018ApJ...864..134R,2019MNRAS.483..152A,2019MNRAS.486.4443F,2019MNRAS.487.5630H,2019PhRvD.100d3010S,2020MNRAS.493.3920F}). 

To first order, field formation predicts that the spins of the BHs are born aligned with the angular momentum of their orbit, while dynamical assembly predicts a random orientation of the spins when the BBH is formed. This fact has been used to distinguish between their formation channels (e.g., \citealt{2020ApJ...892L...8S,2020arXiv200106490S}).

However, in addition to spin, the mass ratio of the BBH is also informative about its assembly. While the field formation scenario has no difficulty in producing low mass-ratio BBH systems, dynamical assembly scenario does. For example, the fraction of the mergers drops by orders of magnitude from those with equal mass ratio to those with $q\approx0.25$, while lower mass ratios are exceedingly rare \citep{2019PhRvD.100d3027R}.

The LIGO/Virgo Scientific Collaboration's recent detection, GW190412 \citep{Abbottetal:2020tl}, comes with properties that are consistent with field formation (the observed low mass ratio) and with dynamical assembly (the observed precession). However, none of the scenarios in their default form can easily explain all of GW190412's characteristics \citep{2020arXiv200506519S}. 

In \citet{2020ApJ...888L...3S}, we suggested that wide hierarchical quadruple systems in the `3+1' configuration (a triple orbited by a fourth body) could account for the merger of a low mass-ratio event with the secondary BH being a mass-gap BH. In this scenario, two neutron stars (NSs) in the innermost orbit of the quadruple system merge due to secular chaotic evolution into a single mass-gap BH. The resulting BH forms a triple system with the remaining two other compact objects (BHs). Due to Lidov-Kozai (LK; \citealt{1962PSS....9..719L,1962AJ.....67..591K}; see, e.g., \citealt{2016ARA&A..54..441N} for a review) oscillations, the mass-gap BH subsequently merges with another BH. A similar configuration could lead to the formation of a BBH merger with masses akin to those of GW190412. Note that, in contrast to a related primordial triple scenario which would require high initial spins, a high spin is naturally explained in our quadruple scenario by the first merger. 

In this paper, we show, based on an extensive suite of numerical integrations, that a hierarchical 3+1 quadruple system is capable of reproducing GW190412-type systems both in terms of its observed mass and spin.
The structure of our paper is as follows: in \S \ref{sec:model}, we explain our model in detail, discuss our methodology, and show some examples of our scenario in action. We present our predicted distributions for the spins and other parameters from population synthesis calculations and compare to GW190412 in \S \ref{sec:result}. We discuss our results in \S \ref{sec:discussion}, and conclude in \S \ref{sec:conclusions}.

\section{Double mergers in 3+1 quadruples}
\label{sec:model}
In this section, we  briefly review our model, discuss our methodology, and give a number of examples.

\subsection{The scenario}
\label{sec:model:scen}
Our scenario is similar to that presented previously in \citet{2020ApJ...888L...3S}. Quadruple systems are common among massive stars: in systems with $\gtrsim 30\,\msun$ primary stars, triples and quadruples are about equally common, and much more common than binaries or singles \citep{2017ApJS..230...15M}. 
Quadruples, which are known to occur in either the 2+2 or 3+1 configuration, can exhibit complex dynamical behavior, which can lead to eccentricity excitation which is more efficient than in equivalent triples \citep{2013MNRAS.435..943P,2015MNRAS.449.4221H,2017MNRAS.466.4107H,2017MNRAS.470.1657H,2018MNRAS.474.3547G,2018MNRAS.478..620H,2019MNRAS.482.2262H,2019MNRAS.483.4060L,2019MNRAS.486.4781F,2020MNRAS.tmp.1284H}.

\begin{figure}
\iftoggle{ApJFigs}{
\includegraphics[width=\linewidth]{configuration}
}{
\includegraphics[width=\linewidth]{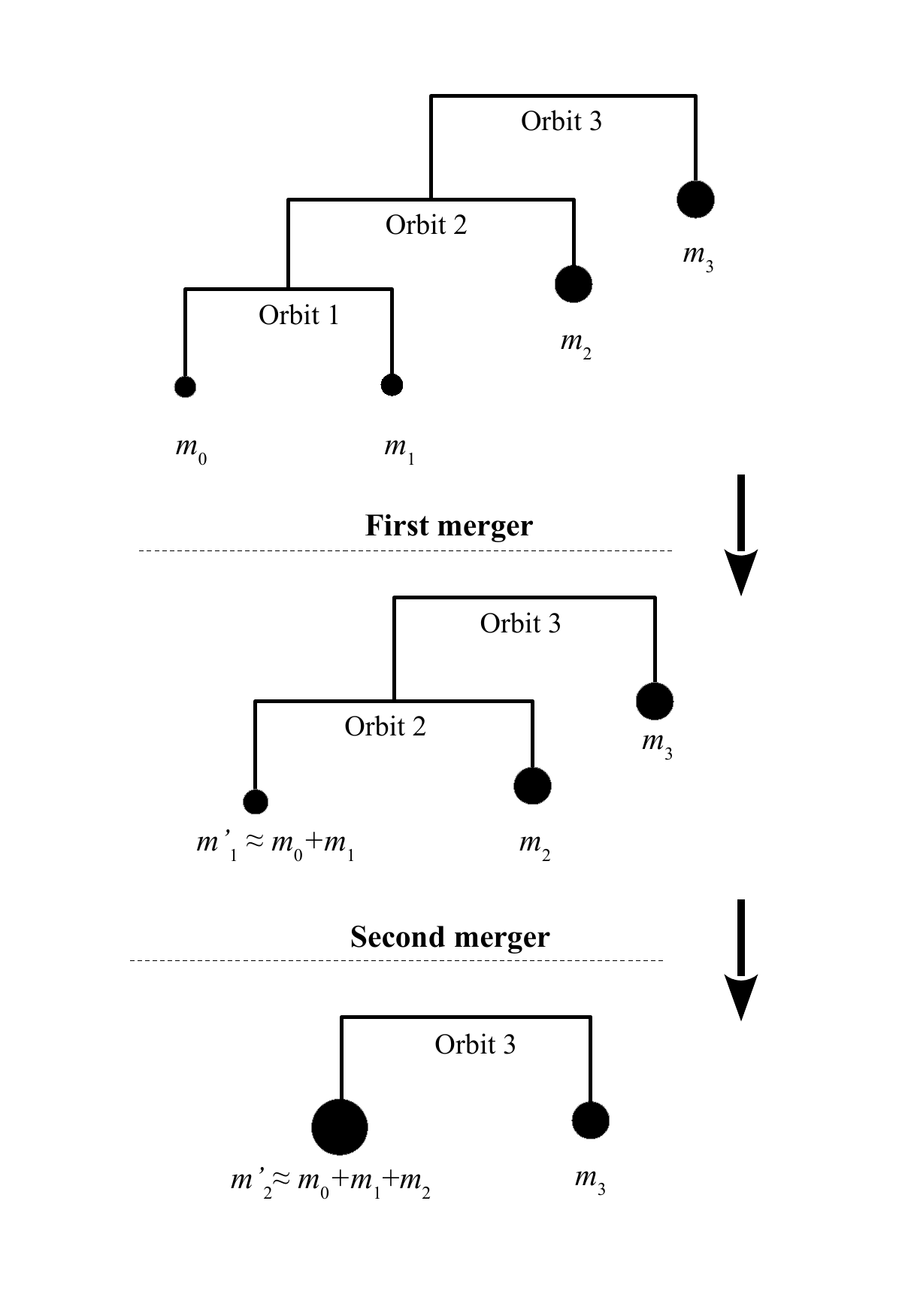}
}
\caption{Sketches of the system in a mobile diagram \citep{1968QJRAS...9..388E}. The system begins as a hierarchical 3+1 quadruple, then merges into a triple (first merger). Finally, the inner binary of the latter merges (second merger). In the paper, any property $x$ of orbit $i$ is denoted as $x_i$.}
\label{fig:config}
\end{figure}

Our proposed channel involves a 3+1 configuration (see \F\,\ref{fig:config}), in which the innermost system (orbit 1) merges due to high eccentricity induced by secular chaotic evolution. The merged body still forms a triple with two other bodies (inner orbit: original orbit 2; outer orbit: original orbit 3), and secular LK evolution and the associated high eccentricities accelerate the merger of the newly-formed inner binary. Finally, a wide binary remains with a GW merger time far exceeding the age of the Universe, and without an outer orbit to excite its eccentricity\footnote{It is possible that a wide binary is driven to high eccentricity and merges due to flybys and/or the Galactic tide (e.g., \citealt{2014ApJ...782...60K,2019ApJ...887L..36M}). We do not consider this possibility for a third merger here. }. 

We assume that a 3+1 quadruple system consisting of four compact objects with semimajor axes of $a_1 \sim 20\,\au$, $a_2 \sim 10^3\,\au$, and $a_3 \sim 10^4\,\au$ can form, despite the existence of pre-compact object evolutionary processes that can potentially prevent its formation. Such processes include orbital expansion due to stellar evolution-induced mass loss causing dynamical instability (e.g., \citealt{2012ApJ...760...99P}), common-envelope (CE) evolution producing a tight compact object binary which is dynamically detached from its distant companions (e.g., \citealt{2013MNRAS.430.2262H}), and natal kicks that can unbind the multiple system (e.g., \citealt{2012MNRAS.424.2914P,2016ComAC...3....6T,2018MNRAS.476.4139H}). Here, we ignore all these complications and leave a self-consistent treatment of pre-compact object evolution to future work. 

The choice of an initially relatively wide innermost orbit, $a_1\sim 20\,\au$, ensures that CE evolution in the innermost orbit is avoided. In order for secular chaotic evolution to effectively excite the innermost orbital eccentricity, the ratio of LK timescales for the orbital pairs (1,2) and (2,3) need to be comparable, i.e., the ratio \citep{2017MNRAS.470.1657H}

\begin{align}
\label{eq:R}
\nonumber \mathcal{R} &\equiv \left ( \frac{a_2^3}{a_1 a_3^2} \right)^{3/2} \left ( \frac{m_0+m_1}{m_0+m_1 + m_2} \right )^{1/2} \frac{m_3}{m_2} \left ( \frac{1-e_2^2}{1-e_3^2} \right)^{3/2} \\
\nonumber & \simeq 0.4 \left ( \frac{a_1}{20 \, \au} \right )^{-3/2} \left ( \frac{a_2}{10^3\,\au} \right )^{9/2} \left ( \frac{a_3}{10^4\,\au} \right )^{-3} \\
&\quad \times \left ( \frac{m_0+m_1}{m_0+m_1 + m_2} \right )^{1/2} \frac{m_3}{m_2} \left ( \frac{1-e_2^2}{1-e_3^2} \right)^{3/2}
\end{align}
needs to satisfy $\mathcal{R} \sim 1$. Here, $e_i$ denotes the (initial) eccentricity of orbit $i$. This consideration motives our choice of initial conditions in the population synthesis calculations (see \S\ref{sec:result:IC} below).

\subsection{Methodology}
\label{sec:model:meth}
We model the secular dynamical evolution of the compact object quadruple using \textsc{SecularMultiple} \citep{2016MNRAS.459.2827H,2018MNRAS.476.4139H,2020MNRAS.tmp.1215H}. The latter code, which is freely available\footnote{\href{https://github.com/hamers/secularmultiple}{https://github.com/hamers/secularmultiple}}, models the evolution based on an expansion of the Hamiltonian of the system in terms of ratios of adjacent orbits, $x$, and orbital averaging. We include binary pair interactions up to and including fifth order in $x$, and binary triplet interactions up to and including third order in $x$. 

In addition to the secular point mass Newtonian evolution, we include post-Newtonian (PN) terms to the 1PN and 2.5PN order, in all orbits. The 1PN terms give rise to orbital in-plane precession, whereas the 2.5PN terms give rise to orbital shrinkage due to GW emission. Here, we ignore PN `interaction' terms that can arise between different orbits (e.g., \citealt{2013ApJ...773..187N,2020arXiv200103654L}). Since we are interested in the spin evolution of the compact objects, we also include the lowest-order spin-orbit coupling terms describing precession of the spins around the orbit, given by \citep{1975PhRvD..12..329B}
\begin{align}
\label{eq:so}
    \frac{\mathrm{d}\unit{S}_i}{\mathrm{d} t } = \frac{2G }{c^2 a_j^3 \left(1-e_j^2\right)^{3/2}} \left ( 1 + \frac{3}{4} \frac{\mci}{m_i} \right ) \ve{L}_j \times \unit{S}_i.
\end{align}
Here, $\ve{S}_i$ (hats denote unit vectors) is the spin angular-momentum vector of body $i$, $m_i$ its mass, $\mci$ the mass of the companion to body $i$ in orbit $j$, and $\ve{L}_j$ is the angular-momentum vector of orbit $j$. The latter has a magnitude given by $L_j =\mu_j \sqrt{G M_j a_j (1-e_j^2)}$, where the total mass is $M_j \equiv m_i+\mci$, and the reduced mass is $\mu_j \equiv m_i \mci/M_j$. Note that \eq~(\ref{eq:so}) implies that the magnitude of $\ve{S}_i$ is conserved. Due to PN spin-orbit coupling, the orbit $\ve{L}_j$ also precesses around the spins; however, the latter effect is negligible if $S_i \ll L_j$, which is satisfied in our case with roughly equal mass-ratio systems (the situation is different when one of the compact objects is supermassive; see, e.g., \citealt{2020arXiv200410205L}). We ignore general relativistic spin-spin coupling, since this is generally only important during the last stages before inspiral, well after the binary has become dynamically decoupled (see below).

We assume that all compact objects are formed with their spin orientations $\unit{S}_i$ aligned with their parent orbit\footnote{Our results would be identical if we had chosen random initial spin orientations, as we discuss later in \S\ref{sec:result:distr:spin}.}, and magnitudes corresponding to half of maximum Kerr rotation, i.e., writing $S_i = \chi_i G m_i^2/c$ with $\chi_i$ the spin parameter ($0\leq \chi_i \leq 1$), we assume that the initial $\chi_i=0.5$. We also take into account different initial spins in post-processing (see \S\ref{sec:result:IC} and \S\ref{sec:result:distr:spin} below). When the two compact objects in the innermost binary merge, we compute the mass (denoted as $m_1'$), spin angular-momentum vector ($\ve{S}_1'$), and recoil velocity of the remnant using the analytic fits of \citet{2010CQGra..27k4006L}. The latter depend on the mass ratio and spin angular-momentum vectors of the two merging compact objects. Depending on the model (see \S\ref{sec:result} below), we subsequently compute the effect of the mass loss and recoil velocity on the newly formed triple system using the routines for external instantaneous perturbations included in \textsc{SecularMultiple} \citep{2018MNRAS.476.4139H}. Below, we indicate properties of the compact object after the first merger with the subscript `1' and a prime (i.e., after first merger, the spin of the merged object is $\ve{S}'_1$, and $\ve{S}_0$ becomes undefined). 

We note that the mass loss and recoil velocity can unbind the triple system, which is particularly the case for unequal mass ratios when the recoil velocity tends to be large. However, we remark that, more generally, large recoil velocities do not necessarily have to impede compact object mergers in quadruples. For example, one of the inner binaries of a 2+2 quadruple could merge, and the imparted recoil velocity could trigger an interaction of the merged compact object with the other inner binary of the quadruple system. The resulting three-body interaction could lead to a second merger event \citep{2020arXiv200211278F}.

In the dynamical integrations, we check for the condition when an orbit becomes decoupled from its secular evolution due to GW emission, i.e., when the timescale for the orbital angular momentum to change by order itself due to secular evolution is ten times longer than the timescale for GW emission to shrink the orbit by order itself (see \citealt{2018ApJ...865....2H}, section 5.1.2). When this condition is satisfied, we stop the integration, since otherwise the integration significantly slows down\footnote{The slowdown is a result of the diverging rate of precession due to the 1PN terms as the orbit shrinks. However, this precession does not affect the evolution since the binary is already decoupled from the outer orbits when we stop the secular integration.}, and after this point in time GW emission completely dominates the evolution. We also remark that, after decoupling, $\ve{L}_j$ no longer changes its direction. The effective spin and precession spin parameters (see \S\ref{sec:result:distr} below) therefore do not change after decoupling (of course, the PN approximation itself breaks down shortly before merger). 

We also check for dynamical instability of the quadruple system during the evolution using the stability criterion of \citet{2001MNRAS.321..398M}, which is applied hierarchically to the (1,2) and (2,3) orbital pairs. In particular, dynamical instability in 3+1 quadruples is often triggered by a secular increase of the eccentricity of orbit 2 (e.g., \citealt{2017MNRAS.466.4107H,2019MNRAS.482.2262H,2020MNRAS.tmp.1284H}).

The simulations of the quadruple systems are run for a duration of $1\,\gyr$. If a merger occurs during this time and the resulting triple is dynamically stable, we continue the integration for an additional $14\,\gyr$. We restrict the first integration duration since the integration of the quadruple system is relatively computationally expensive. From that point of view, our merger fractions should be considered to be lower limits.

\subsection{Examples}
\label{sec:model:ex}

\begin{figure*}
\iftoggle{ApJFigs}{
\includegraphics[width=0.48\linewidth]{system_76849}
\includegraphics[width=0.48\linewidth]{system_72571}
}{
\includegraphics[width=0.48\linewidth]{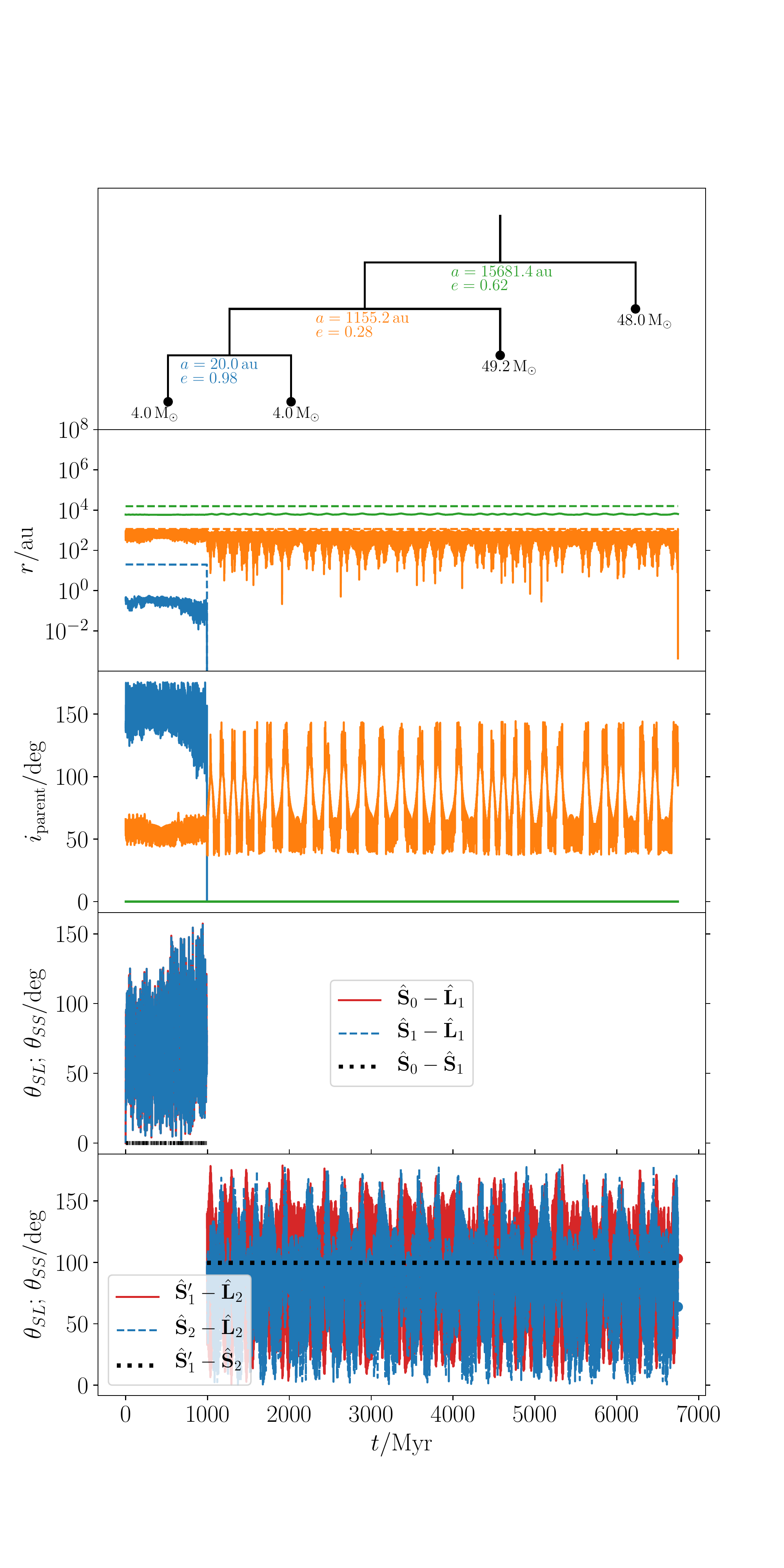}
\includegraphics[width=0.48\linewidth]{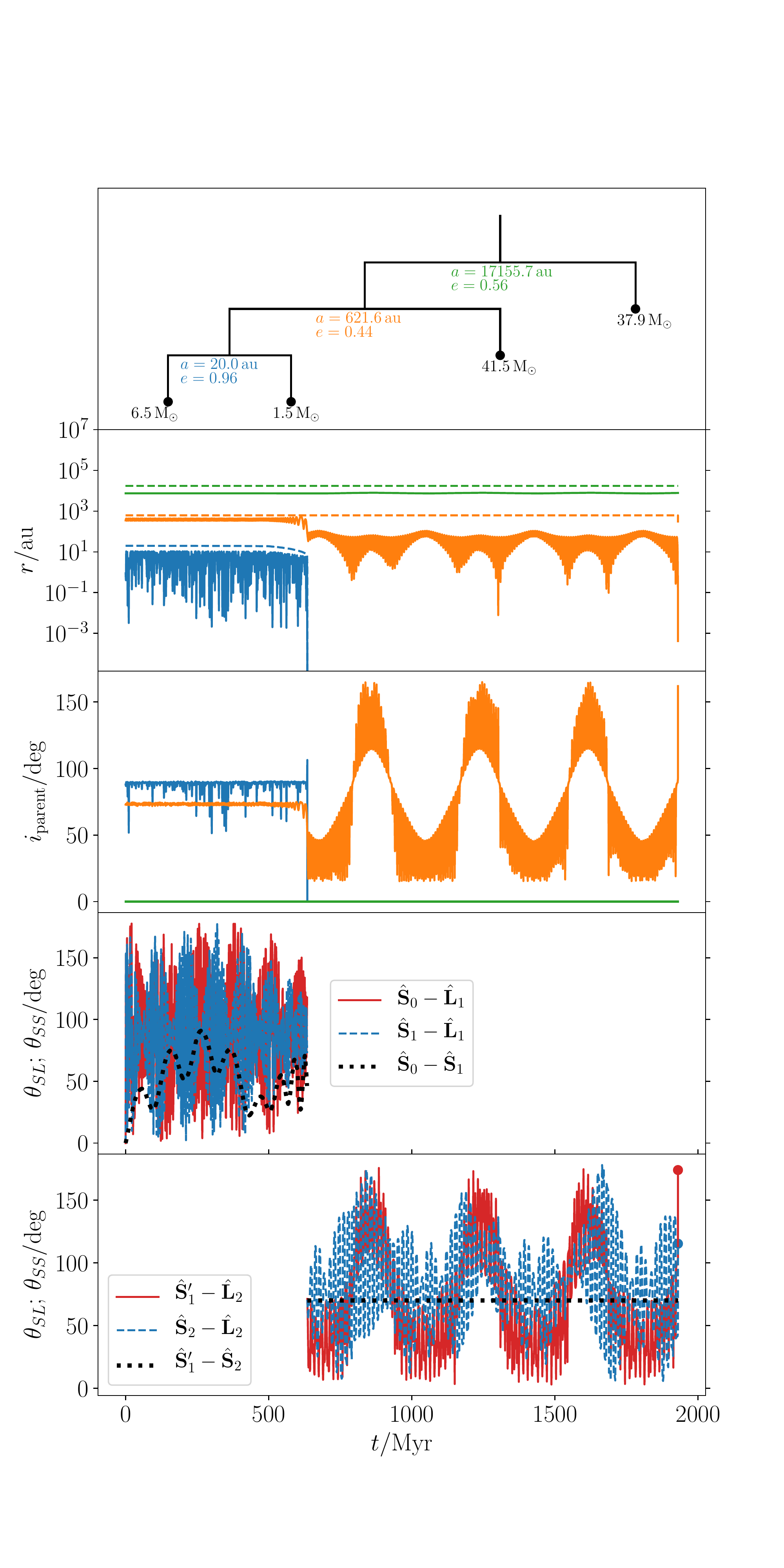}
}
\caption{ Two examples in which double mergers occur in 3+1 quadruple systems. Left-hand panels: $m_0=m_1$; right-hand panels: $m_0>m_1$. The top panels show the initial configuration of the system in Mobile diagrams \citep{1968QJRAS...9..388E}, with some of the initial orbital parameters indicated. The second panels from the top show the orbital separations (dashed: semimajor axes; solid: periapsis distances, $r_{\mathrm{p},j} = a_j[1-e_j]$); the colors correspond to the colors in the top panels. The third panels from the top show the mutual inclinations between the parent orbits. The bottom two panels show various spin-orbit and spin-spin angles. The fourth rows apply to before the first merger; they show the spin-orbit angles with respect to the innermost orbit, orbit 1: $\unit{S}_0$-$\unit{L}_1$ (solid red lines) and $\unit{S}_1$-$\unit{L}_1$ (blue dashed lines). Also shown is the spin-spin angle for bodies 0 and 1 ($\unit{S}_0$-$\unit{S}_1$; black dotted lines). The fifth rows apply to after the first merger; they show the spin-orbit angles for the merged object, body 1', and body 2 with respect to the now inner orbit of the triple, orbit 2: $\unit{S}'_1$-$\unit{L}_2$ (solid red lines) and $\unit{S}_2$-$\unit{L}_2$ (blue dashed lines). Also shown is the spin-spin angle for bodies 1' and 2 ($\unit{S}'_1$-$\unit{S}_2$; black dotted lines). The red and blue dots in the bottom panels indicate the spin-orbit angles of bodies 1' and 2 at the moment of the second merger. }
\label{fig:ex}
\end{figure*}
We illustrate our scenario with two examples in \F~\ref{fig:ex}. In the left-hand panels (example taken from Model D as defined below in \S\ref{sec:result:IC}), the innermost two masses are $m_0=m_1=4\,\msun$ (two low-mass BHs). In the right-hand panels (from Model C as defined in \S\ref{sec:result:IC}), the innermost two masses are unequal, $m_0=6.5\,\msun$ (BH), and $m_1=1.5\,\msun$ (NS). The top panels show the initial configuration of the system in Mobile diagrams \citep{1968QJRAS...9..388E}, with the initial orbital parameters indicated. The second panels from the top show the orbital separations, the third panels the mutual inclinations, and the bottom panels show various spin-orbit and spin-spin angles. 

In both examples, the innermost orbit is initially driven to high eccentricity through secular chaotic evolution. After the first merger, the disappearance of the innermost orbit changes the character of the secular oscillations of orbit 2 (now the inner orbit of the triple). In these examples in which a double merger occurred, the mutual inclination between binaries 2 and 3 (the inner and outer orbits of the triple) after first merger is close to $90^\circ$. Also, the inner orbit of the triple is wide, such that 1PN precession is ineffective at quenching LK oscillations (e.g., \citealt{2002ApJ...578..775B,2003ApJ...598..419W,2011ApJ...741...82T,2015MNRAS.447..747L}). A second merger in the triple system, which requires an extremely high eccentricity ($1-e_2 \sim 10^{-7}$, see also \S\ref{sec:result:distr} below) can therefore be achieved, in particular when the orbital orientation switches between prograde and retrograde, or vice versa. 

The bottom two panels in \F~\ref{fig:ex} show the spin-orbit and spin-spin angles. In the equal-mass case, the innermost two compact objects precess around $\unit{L}_1$ at the same rate (see \eq~\ref{eq:so}). Therefore, the spin-spin angle associated with $\unit{S}_0$ and $\unit{S}_1$ remains $0^\circ$ (see the black dotted line). Also, the spin-orbit angles between bodies 0 and 1 are identical. In the unequal-mass case, $\unit{S}_0$ and $\unit{S}_1$ precess at a differential rate, producing spin-spin and spin-orbit misalignment. After the first merger (fifth rows in \F~\ref{fig:ex}), the angle between the now-merged $\unit{S}'_1$ and $\unit{S}_2$ remains constant, both in the equal-mass case, and in the unequal-mass case. In both cases, the compact objects have unequal masses, but owing to the large separation of the (now inner) orbit, spin-orbit coupling is unimportant. Also, note that $\unit{S}'_1$ and $\unit{S}_2$ are generally not mutually aligned. 

The spin-orbit angles associated with $\unit{S}'_1$ and $\unit{S}_2$ and $\unit{L}_2$ are therefore solely driven by the changing direction of $\unit{L}_2$, which is caused by the secular torque of orbit 3. Consequently, the spin-orbit angles for $\unit{S}'_1$ and $\unit{S}_2$ at the moment of second merger are essentially random (see also \F~\ref{fig:thetas} below).

\section{Results}
\label{sec:result}
\subsection{Initial conditions}
\label{sec:result:IC}
We use Monte Carlo sampling to explore a restricted parameter space of compact object quadruples, which is informed by the condition \eq~(\ref{eq:R}). As discussed in \S\ref{sec:model:scen}, we ignore any pre-compact object evolution, and start the dynamical integrations with four compact objects. The initial conditions of our systems are highly uncertain; here, we assume simple and idealised distributions. 

We set the masses $m_0$ and $m_1$ to fixed values depending on the model, whereas $m_2$ and $m_3$ are sampled between $20\,\msun$ and $50\,\msun$ assuming uniform distributions. The innermost semimajor axis is set to $a_1=20\,\au$ (to avoid prior CE evolution), whereas $a_2$ and $a_3$ are sampled from distributions flat in their log values, with $500\,\au<a_2<2000\,\au$, and $1.4\times10^4\,\au<a_3<2.0 \times 10^4\,\au$. The orbital eccentricities are sampled from thermal distributions, $\mathrm{d}N/\mathrm{d}e_j\propto e_j$, with $0.01<e_j<0.99$. The orbital angles (inclination, argument of periapsis, and longitude of the ascending node) are sampled according to isotropic orbital orientations (for all orbits). We reject systems that are initially dynamically unstable according to the stability criterion of \citet{2001MNRAS.321..398M}, which is applied hierarchically to the (1,2) and (2,3) orbital pairs. Note that the initial inner orbit periapsis distance is always larger than $0.2\,\au$ since $e_1<0.99$.  

\begin{figure}
\iftoggle{ApJFigs}{
\includegraphics[width=1\linewidth]{test02_N_MC_100000_mode_chi_i_0_chi_1p}
}{
\includegraphics[width=1\linewidth]{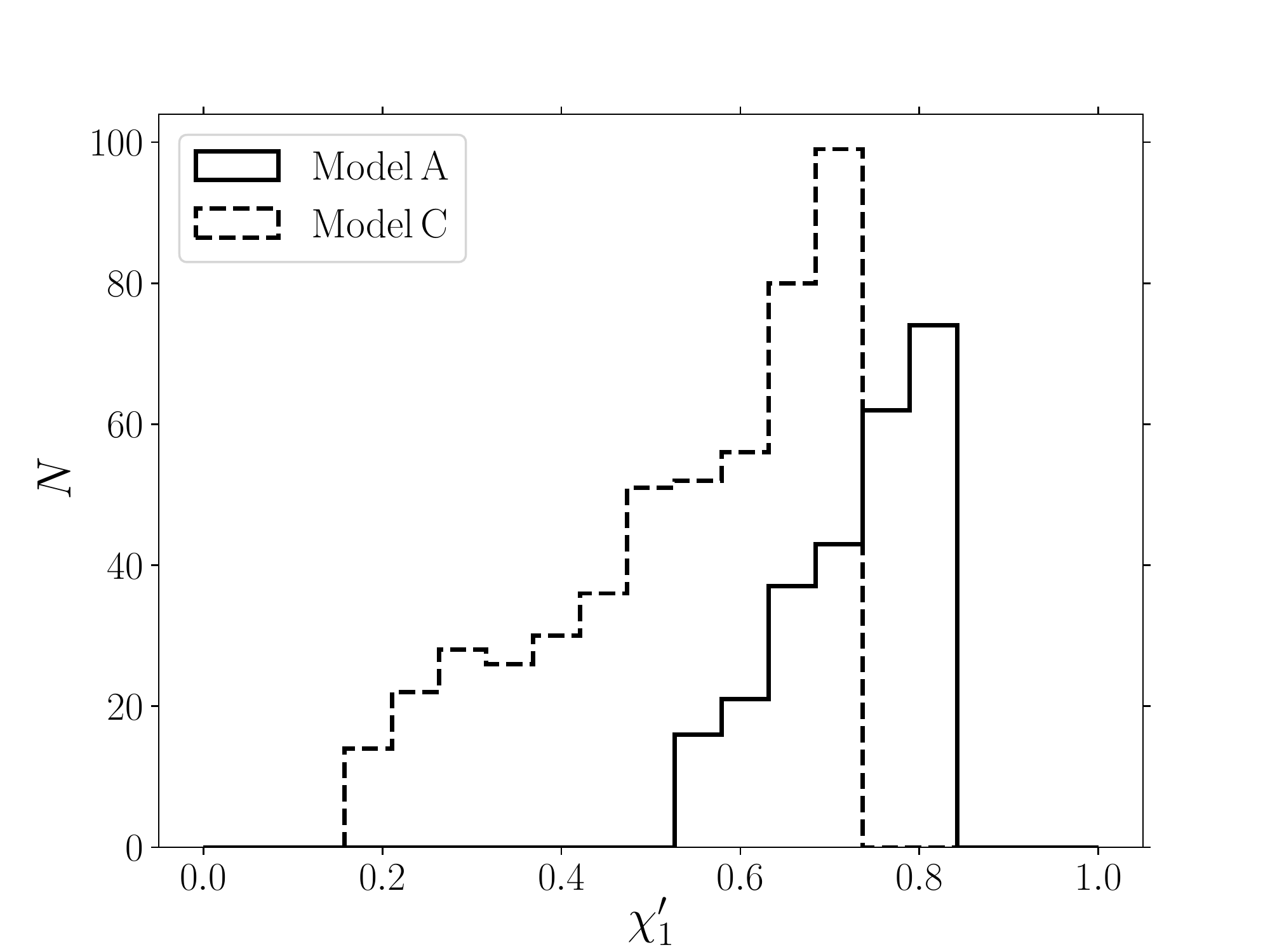}
}
\caption{ Distributions of the spin of the merged object, $\chi'_1$, for model A (solid line), and model C (dashed line). The initial spins in these cases were $\chi_0=\chi_1=0.5$. }
\label{fig:chi1p}
\end{figure}

The initial spin angular momenta of all bodies are assumed to be aligned with their parent orbital momenta. Their magnitudes are set corresponding to $\chi_i=0.5$ (see also \S\ref{sec:model:meth}). We also take into account different values of $\chi_2$, which does not change in the simulations, in post-processing (see \S\ref{sec:result:distr:spin} below). Regarding the impact of different $\chi_i$ on the spin of the merged object, $\chi_1'$, we note that, assuming spins aligned with the orbit, two non-spinning equal-mass BHs will make a BH with $\chi'\approx0.69$, while if the two BHs had an initial $\chi_i=0.5$, the final remnant has a spin of $\chi'\approx0.83$ \citep{2010CQGra..27k4006L}. Similarly for the case of a $6.5+1.5\,\msun$ system, the assumption of $\chi_i=0.5$ leads to a final spin of $\chi'\approx0.72$ while non-spinning BHs will result in a final spin of $\chi'\approx0.45$. These ranges fall well within the distributions of $\chi_1'$ when a fixed $\chi_i=0.5$ is assumed (see \F~\ref{fig:chi1p}), indicating that $\chi_1'$ is insensitive to the assumed initial value of $\chi_i$. Moreover, our spin parameter distributions are mostly affected by $\chi_2$, and the impact of varying $\chi_2$ is explored in Appendix~\ref{app:spin}.

\begin{table}
\centering
\begin{tabular}{cccccc}
\toprule
 Model & $m_0/\msun$ & $m_1/\msun$ & $\unit{S}$-$\unit{L}$ & Recoil & $\fdm\,(\%)$ \\
\midrule
A & 4.0 & 4.0 & \cmark & \xmark & $0.253 \pm 0.016$ \\
B & 4.0 & 4.0 & \xmark & \xmark & $0.204 \pm 0.014$ \\
C & 6.5 & 1.5 & \cmark & \xmark & $0.494 \pm 0.022$ \\
D & 4.0 & 4.0 & \cmark & \cmark & $0.193 \pm 0.014$ \\
E & 6.5 & 1.5 & \cmark & \cmark & 0 \\
\bottomrule
\end{tabular}
\caption{Summary of the five different models for the masses $m_0$ and $m_1$, whether or not spin-orbit terms were included (\eq~\ref{eq:so}; column `$\unit{S}$-$\unit{L}$'), and whether or not recoil and mass loss after the first merger was taken into account (column `Recoil'). The right-most column shows the double merger fraction ($f_\mathrm{dm}$, in per cent), together with the Poisson-based error. }
\label{table:models}
\end{table}

We adopt five different models (labeled A through E), in which we investigate the importance of the spin-orbit terms and the recoil velocity after the first merger, and the impact of the mass ratio in the innermost binary. The models are summarized in Table~\ref{table:models}. We sample $N_\mathrm{MC}=10^5$ systems for each model (giving a total of $5\times10^5$ systems). Note that, when `Recoil' terms were not included, we also excluded the effects of instantaneous mass loss after the first merger on the triple system. However, these effects are typically small since the mass loss after the first merger event is small (up to a few per cent). 

The double merger fractions $\fdm$ in our simulations are presented (in per cent) in the right-most column of Table~\ref{table:models}. The fractions are generally low, i.e., between 0.2 and 0.5 per cent. Including the spin-orbit terms (models A vs. B) does not affect the rates beyond statistical significance, which is to be expected since, in our simulations, the spins can only precess around the orbits, and the spins themselves do not affect the dynamical evolution. We remark that, in practice, the merger fractions are different nevertheless. This is a numerical artefact, and can be attributed to the chaotic nature of the system: including the spin-orbit terms affects the internal time steps taken by the ordinary differential equation integrator used in \textsc{SecularMultiple}. Since the system is chaotic, this can lead to different outcomes for the same initial conditions. 

Interestingly, when recoil is not taken into account, choosing unequal masses in the innermost system boosts the double merger fraction by a factor of $\sim 2$ (models A vs. C). However, when recoil is taken into account (models C vs. E), double mergers no longer occur at all in the unequal mass model. This can be attributed to the large recoil velocity (typically $200\,\kms$, with a tail extending to $\sim 500\,\kms$). In the equal-mass case, however, (models A vs. D), the recoil velocity is zero and the only effect of including the `Recoil' terms is the small mass loss after the first merger; the effect on $\fdm$ is relatively small.

\subsection{Distributions}
\label{sec:result:distr}

\begin{figure}
\iftoggle{ApJFigs}{
{\includegraphics[width=1\linewidth,trim = 0mm 0mm 0mm 0mm]{test02_N_MC_100000_mode_chi_i_0_chi_eff}}
}{
{\includegraphics[width=1\linewidth,trim = 0mm 0mm 0mm 0mm]{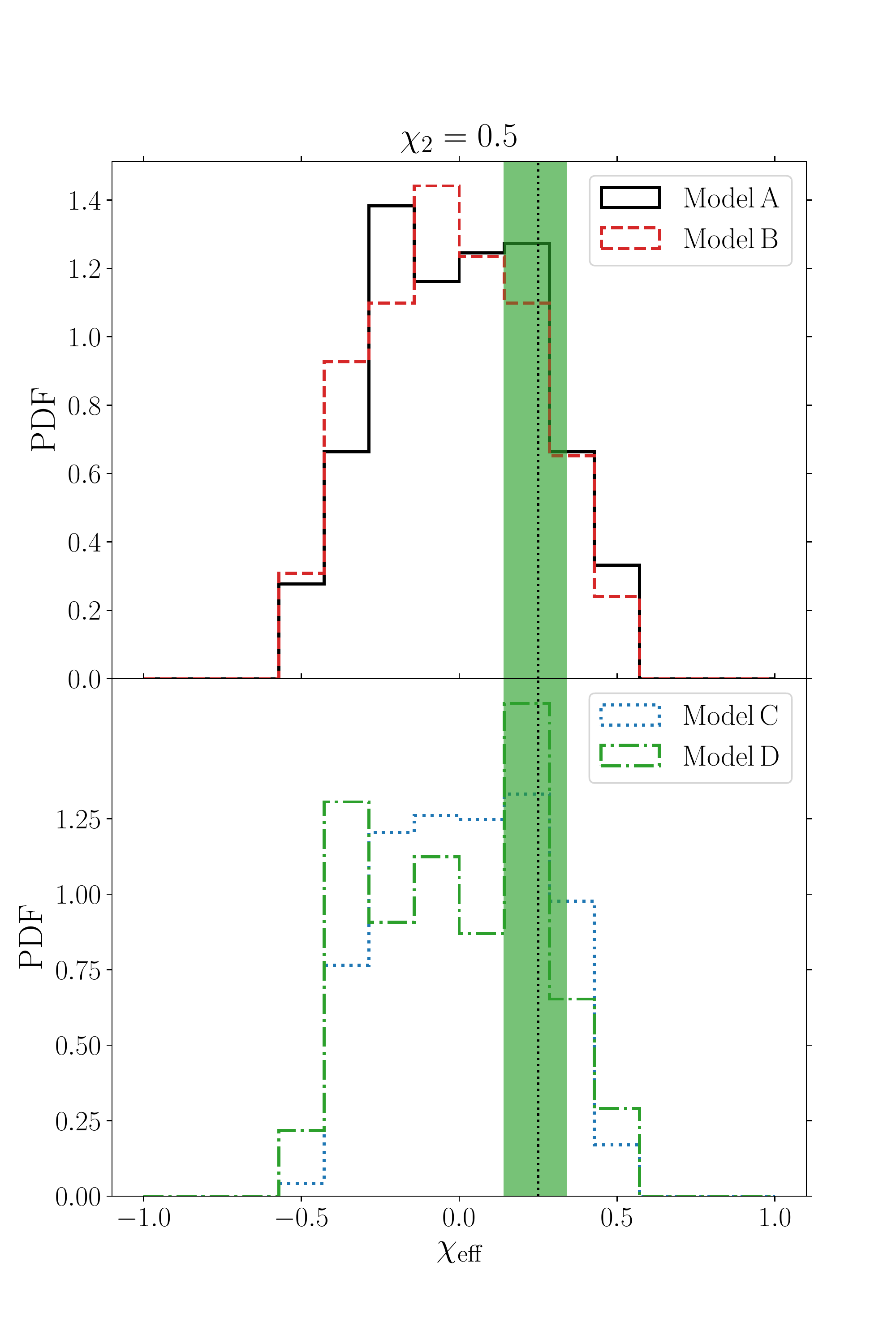}}
}
\caption{ Distributions of $\chie$ (see \eq~\ref{eq:chi_e}) for the double mergers in our simulations (note that there are none in model E). The top (bottom) panels show distributions for models A and B (C and D), with line styles and colors as indicated in the legend. The LIGO/Virgo value for GW190412, $\chie=0.25_{-0.11}^{+0.09}$ \citep{Abbottetal:2020tl}, is shown with the dotted black lines and green shaded regions. }
\label{fig:chi_eff}
\end{figure}

\subsubsection{Spins}
\label{sec:result:distr:spin}
In this section, we present several spin-related distributions for the components $m'_1$ (the merged compact object after first merger) and $m_2$ relative to their orbit $\unit{L}_2$, at the time of the second merger. We first consider the effective spin parameter $\chie$, i.e., the mass-weighted spin onto orbit $\unit{L}_2$ which is defined according to
\begin{align}
\label{eq:chi_e}
    \chie \equiv \frac{\chi'_1 m'_1 \left(\unit{S}'_1\cdot \unit{L}_2\right) + \chi_2 m_2 \left(\unit{S}_2\cdot \unit{L}_2 \right )}{m'_1+m_2}.
\end{align}
Here, we determine $\chie$ from the simulations for the double merger systems when the inner binary of the triple has become dynamically decoupled from $m_3$ (see \S\ref{sec:model:meth}). As described in \S\ref{sec:model:meth}, the parameters $\chi_1'$ and  $m_1'$ of the compact object after first merger are determined using the fits from \citet{2010CQGra..27k4006L}. Note that, after decoupling, $\unit{L}_2$ is constant, but the spins $\unit{S}'_1$ and $\unit{S}_2$ still precess around $\unit{L}_2$ according to the spin-orbit terms, \eq~(\ref{eq:so}). However, this does not affect the value of $\chie$, since \eq~(\ref{eq:so}) conserves the spin-orbit projections, $\unit{S}'_1\cdot \unit{L}_2$, and $\unit{S}_2\cdot\unit{L}_2$. 

We show the distributions of $\chie$ in \F~\ref{fig:chi_eff} for the double mergers in our simulations (note that there are none in model E). All of our models produce distributions of $\chie$ that are centered and approximately symmetric around zero, with a tail extending to approximately $\chie \pm 0.5$. Since $m_2 \gg m'_1$, the latter value is mostly dictated by $\chi_2$, which in our models was set to $\chi_2=0.5$ (note that $\chi_2$ does not change in the simulations). All models are consistent with the LIGO/Virgo value for GW190412 \citep{Abbottetal:2020tl}, $\chie=0.25_{-0.11}^{+0.09}$ (indicated in the figure with the dotted black lines and green shaded regions). 

\begin{figure}
\iftoggle{ApJFigs}{
\includegraphics[width=1.\linewidth]{test02_model_0_N_MC_100000_mode_chi_i_0_theta_sls}
}{
\includegraphics[width=1.\linewidth]{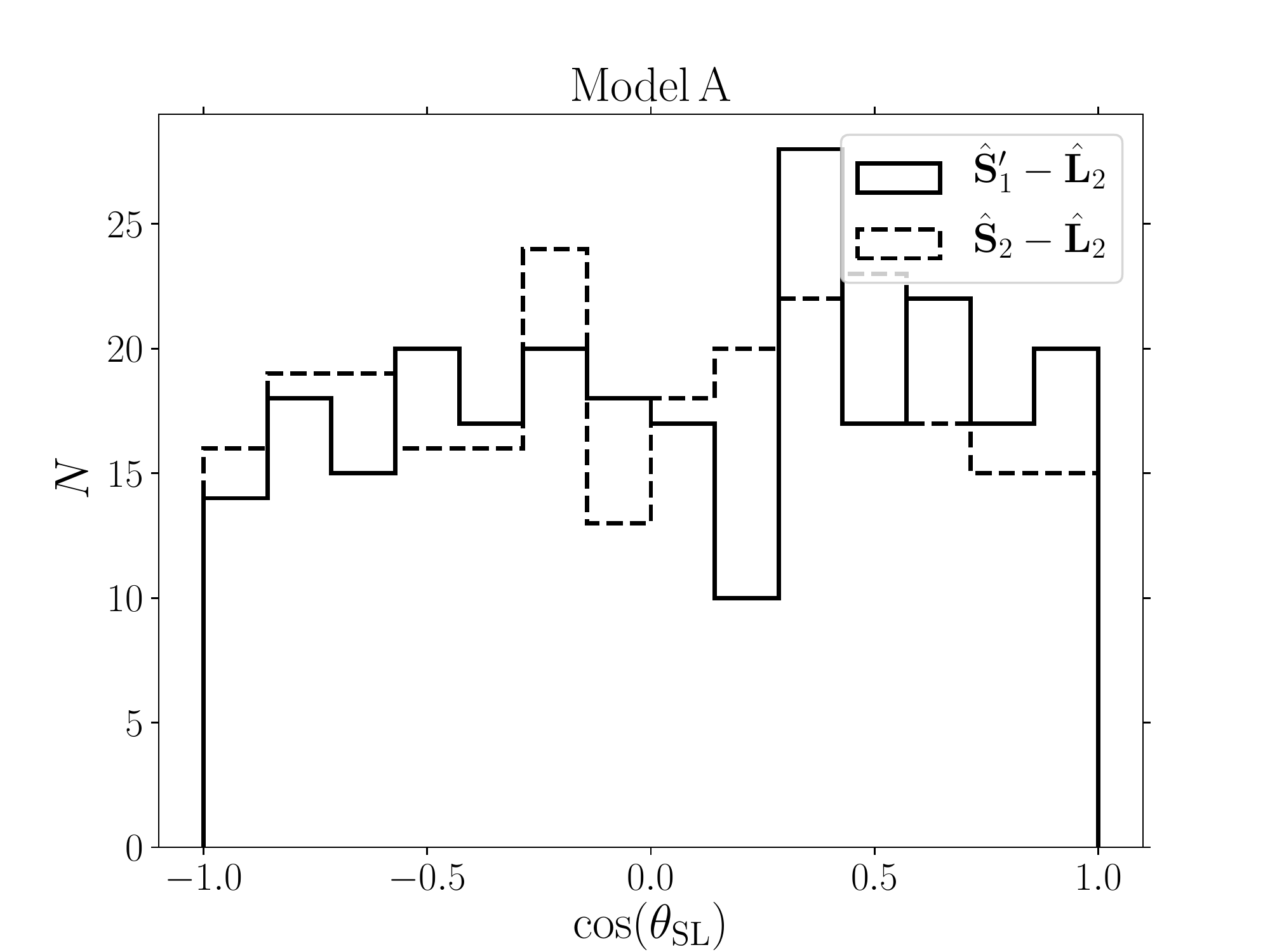}
}
\caption{ Spin-orbit distributions of the cosines of the angles between $\unit{S}'_1$ and $\unit{L}_2$ (solid line), and $\unit{S}_2$ and $\unit{L}_2$ (dashed line), by the time of the second merger. These distributions apply to model A; results are qualitatively similar for the other models. }
\label{fig:thetas}
\end{figure}

The fact that the distributions of $\chie$ are centered around zero can be explained from the spin-orbit distributions. \F~\ref{fig:thetas} shows the spin-orbit distributions for the angles between $\unit{S}'_1$ and $\unit{L}_2$, and $\unit{S}_2$ and $\unit{L}_2$, by the time of the second merger (for model A; other models give qualitatively similar results). These distributions are approximately flat in their cosines, indicating isotropic orientations between spins and orbit. This can be understood by noting that spin-orbit coupling is only important before the first merger; after the first merger, the wide inner orbit of the triple implies that spin-orbit coupling is unimportant (see also the examples in \S\ref{sec:model:ex}). The spin-orbit angles are therefore driven solely by the changing direction of $\unit{L}_2$ by the secular torque of $m_3$, which leads to a random orientation by the time of second merger. This is similar to what has been found in studies of the spins of merging compact objects in isolated triples (e.g., \citealt{2017ApJ...846L..11L,2018MNRAS.480L..58A,2020MNRAS.493.3920F}). The randomisation of the spin-orbit angles by secular evolution by the time of the second merger also implies that the distributions of the spin-orbit angles are independent of the initial assumed orientations.  

Next, we consider the in-plane components of the spins with the orbit $\unit{L}_2$. Owing to its asymmetric masses, GW190412 shows stronger contributions from higher-multipole GWs, which gives constraints on the in-plane components of the spins, 
\begin{align}
\unit{S}_{i,\perp}=\unit{S}_i \left[ 1 - \left( \unit{S}_i \cdot \unit{L}_2 \right ) \right ],    
\end{align}
through the parameter $\chip$ which is defined according to \citep{Abbottetal:2020tl}
\begin{align}
\label{eq:chi_p}
    \chip \equiv \mathrm{max} \left [ \kappa \chi'_1 \left|\left|\unit{S}'_{1,\perp}\right|\right|,  \chi_2 \left|\left|\unit{S}_{2,\perp}\right |\right |  \right ].
\end{align}
Here, $\kappa = q(4q+3)/(4+3q)$ with $q=m'_1/m_2$. Note that in \citet{Abbottetal:2020tl}, $\chip$ was defined with $m_2$ being the less massive component, whereas in our case, $m_2$ is the more massive component. Also, note that the spin-orbit terms (\eq~\ref{eq:so}) do not change $\chip$ after decoupling since $\left|\left|\unit{S}_{i,\perp}\right|\right|=1- \left( \unit{S}_i \cdot \unit{L}_2 \right )$.

\begin{figure}
\iftoggle{ApJFigs}{
\includegraphics[width=1\linewidth]{test02_N_MC_100000_mode_chi_i_0_chi_p}
}{
\includegraphics[width=1\linewidth]{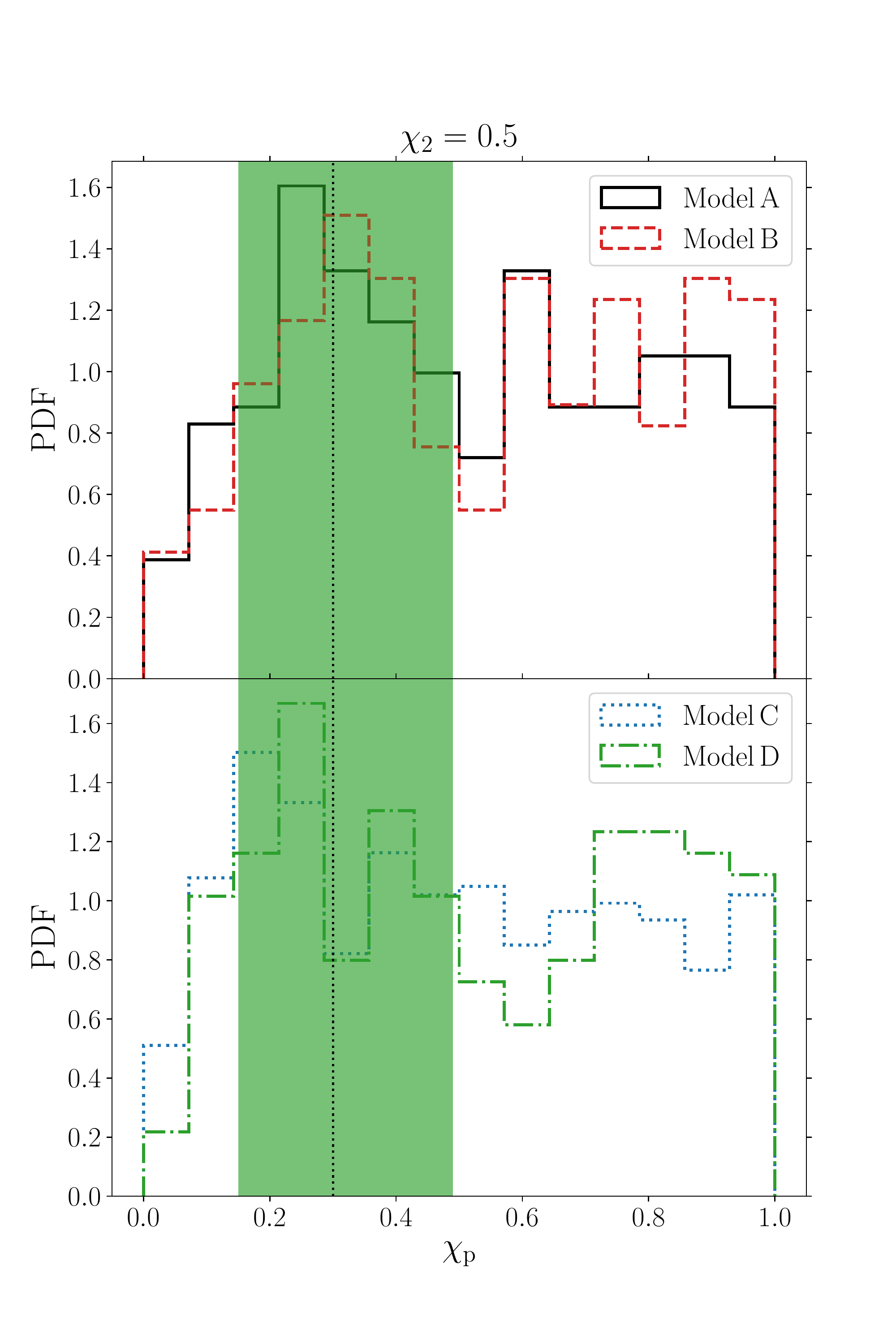}
}
\caption{ Distributions of $\chip$ (\eq~\ref{eq:chi_p}) for the double mergers in our simulations (note that there are none in model E). The top (bottom) panels show distributions for models A and B (C and D), with line styles and colors as indicated in the legend. The LIGO/Virgo value for GW190412, $\chip=0.30_{-0.15}^{+0.19}$ \citep{Abbottetal:2020tl}, is shown with the dotted black lines and green shaded regions. }
\label{fig:chi_p}
\end{figure}

The distributions of $\chip$ for the different models in our simulations are shown in \F~\ref{fig:chi_p}. The different models give qualitatively similar distributions; $\chip$ is broadly distributed between 0 and 1, with a slight preference for $\chip\sim 0.2$. Interestingly, the latter is consistent with the LIGO/Virgo value for GW190412, $\chip=0.30_{-0.15}^{+0.19}$ \citep{Abbottetal:2020tl}.

In the above, $\chi_2$ was set to $\chi_2=0.5$. The birth spins $\chi_i$ of BHs are uncertain, and some BHs might be born with low spins (e.g., \citealt{2019MNRAS.485.3661F}). For the double mergers in our scenario, the merged compact object tends to have a relatively high spin of $\chi'_1\sim 0.7$, largely independent of the initial $\chi_0$ and $\chi_1$ (e.g., \citealt{2010CQGra..27k4006L}). However, the spin $\chi_2$ could be significantly different than the assumed $\chi_2=0.5$. Therefore, we also computed the distributions of $\chie$ and $\chip$ in post-processing by making different assumptions on $\chi_2$ (note that the spin-orbit terms, \eq~\ref{eq:so}, do not affect the magnitudes of the spins). In particular we either assume $\chi_2=0.1$, or sample $\chi_2$ from a uniform distribution with $0<\chi_2<1$. The resulting distributions of $\chie$ and $\chip$ are shown in \Fs~\ref{fig:chi_eff_app} and \ref{fig:chi_p_app} in Appendix~\ref{app:spin}. 

With these different $\chi_2$, the distributions of $\chie$ (\F~\ref{fig:chi_eff_app}) are still centered and symmetric around 0, but the extent of the distribution is affected. Since body 2 dominates $\chie$, when $\chi_2=0.1$, the distribution of $\chie$ is more confined, within $\chip=\pm0.25$, and is only marginally consistent with GW190412. With the uniform distribution of $\chi_2$, the distribution extends to approximately $\chie=\pm1$, and is consistent with GW190412. 

The distribution of $\chip$ (\F~\ref{fig:chi_p_app}) with $\chi_2=0.1$ is strongly concentrated around $\chip=0.2$, but still consistent with GW190412. With the uniform distribution of $\chi_2$, the distribution of $\chip$ is more broad, but still peaks around $\chip=0.2$.

\subsubsection{Orbital properties}
\label{sec:result:distr:orbit}
In the remaining sections (\S\ref{sec:result:distr:orbit}-\S\ref{sec:result:distr:time}), we discuss some of the orbital, mass, and merger time distributions. These distributions are typically qualitatively not strongly dependent on the model; here, we show distributions from model A. 

\begin{figure}
\iftoggle{ApJFigs}{
\includegraphics[width=\linewidth]{test02_model_0_N_MC_100000_mode_chi_i_0_GW_a0s}
\includegraphics[width=\linewidth]{test02_model_0_N_MC_100000_mode_chi_i_0_GW_e0s}
}{
\includegraphics[width=\linewidth]{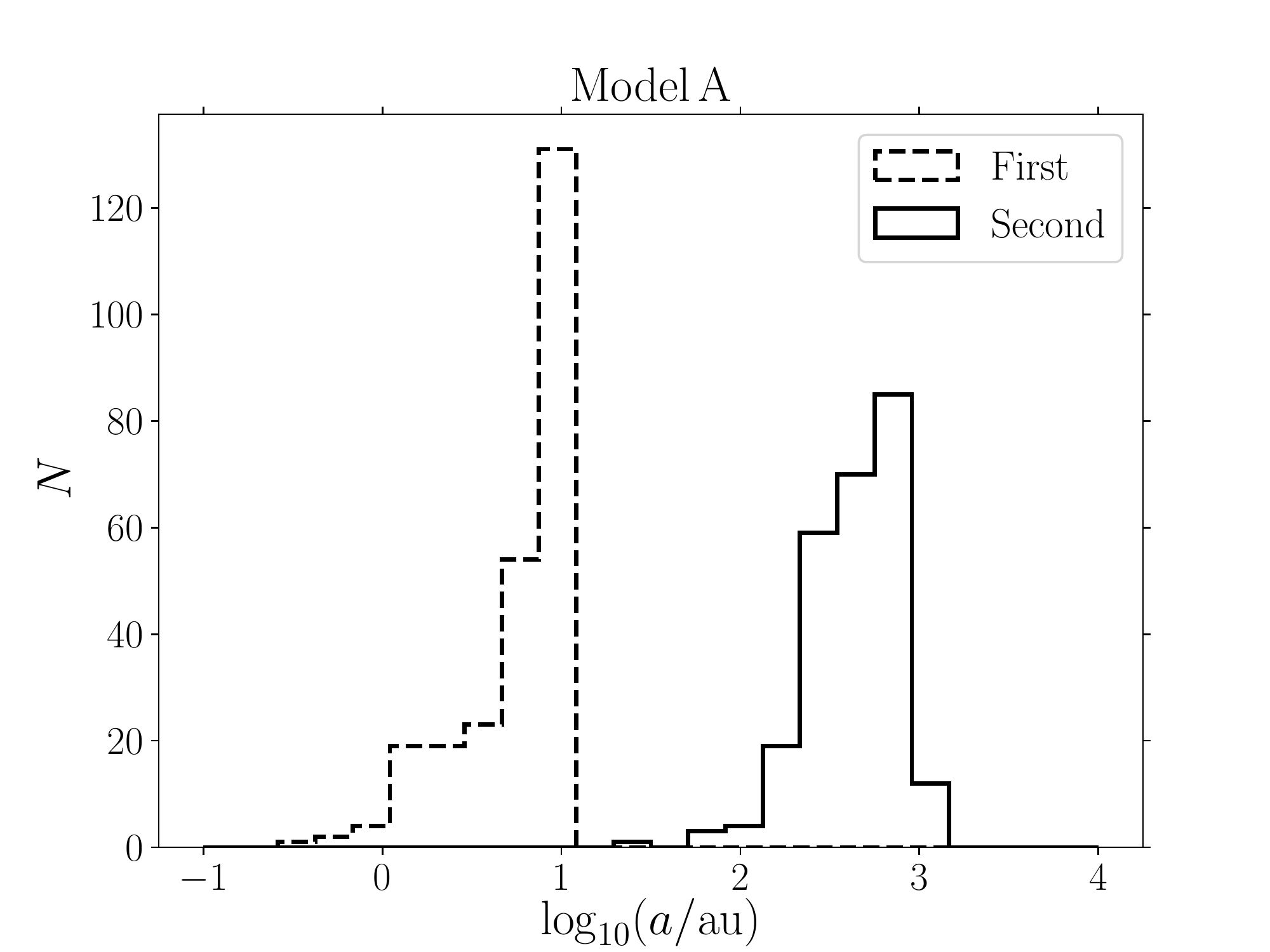}
\includegraphics[width=\linewidth]{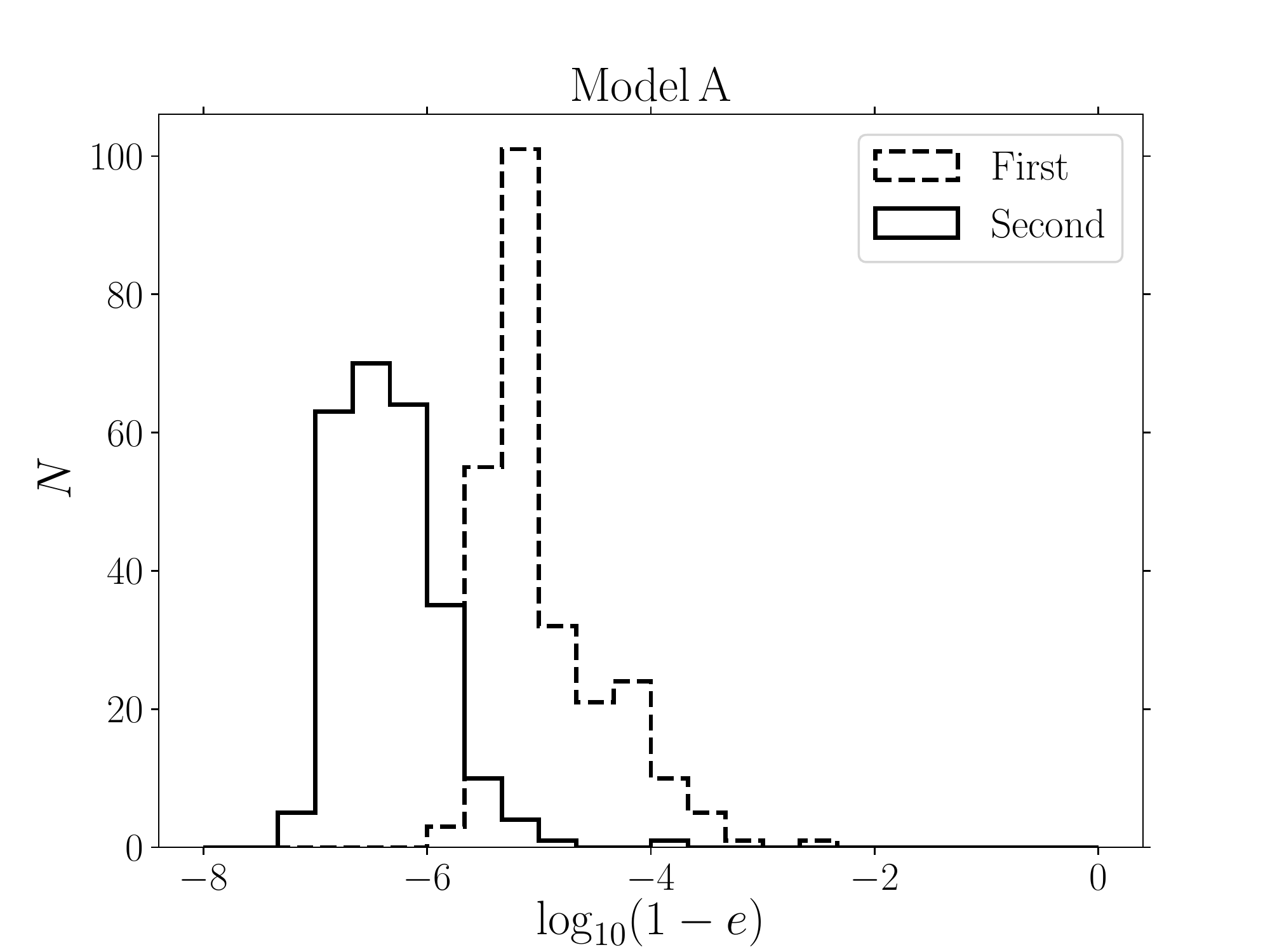}
}
\caption{Distributions of the semimajor axes (top panel) and eccentricities (bottom panel) of the merging orbits at the moment of decoupling. Dashed and solid lines correspond to the first and second mergers, respectively. }
\label{fig:GWae0}
\end{figure}

\F~\ref{fig:GWae0} shows the distributions of the semimajor axes and eccentricities of the merging orbit at the moment of decoupling. Dashed and solid lines correspond to the first and second mergers, respectively. For the first merger, $a_1$ is peaked around $20\,\au$, reflecting the initial value $a_1=20\,\au$. A tail exists towards smaller values, which is due to orbital shrinkage due to GW emission before decoupling (an example of this can be seen in the right-hand panels of \F~\ref{fig:ex}). For the second merger, $a_2$ is distributed between $\sim 10^2$ and $10^3\,\au$, mostly reflecting the initial conditions ($500\,\au<a_2<2000\,\au$, see \S\ref{sec:result:distr}). In some cases, GW emission resulted in significant shrinkage before decoupling. 

Since the merging orbits in our simulations are wide, the eccentricities at decoupling (bottom panel of \F~\ref{fig:GWae0}) are necessarily extremely high. The eccentricity at the first merger ($\sim 1-10^{-5}$) is lower than at the second merger ($\sim1-10^{-7}$), since the orbit is wider in the latter case (by roughly two orders of magnitude).

\begin{figure}
\iftoggle{ApJFigs}{
\includegraphics[width=\linewidth]{test02_model_0_N_MC_100000_mode_chi_i_0_es_LIGO_band}
}{
\includegraphics[width=\linewidth]{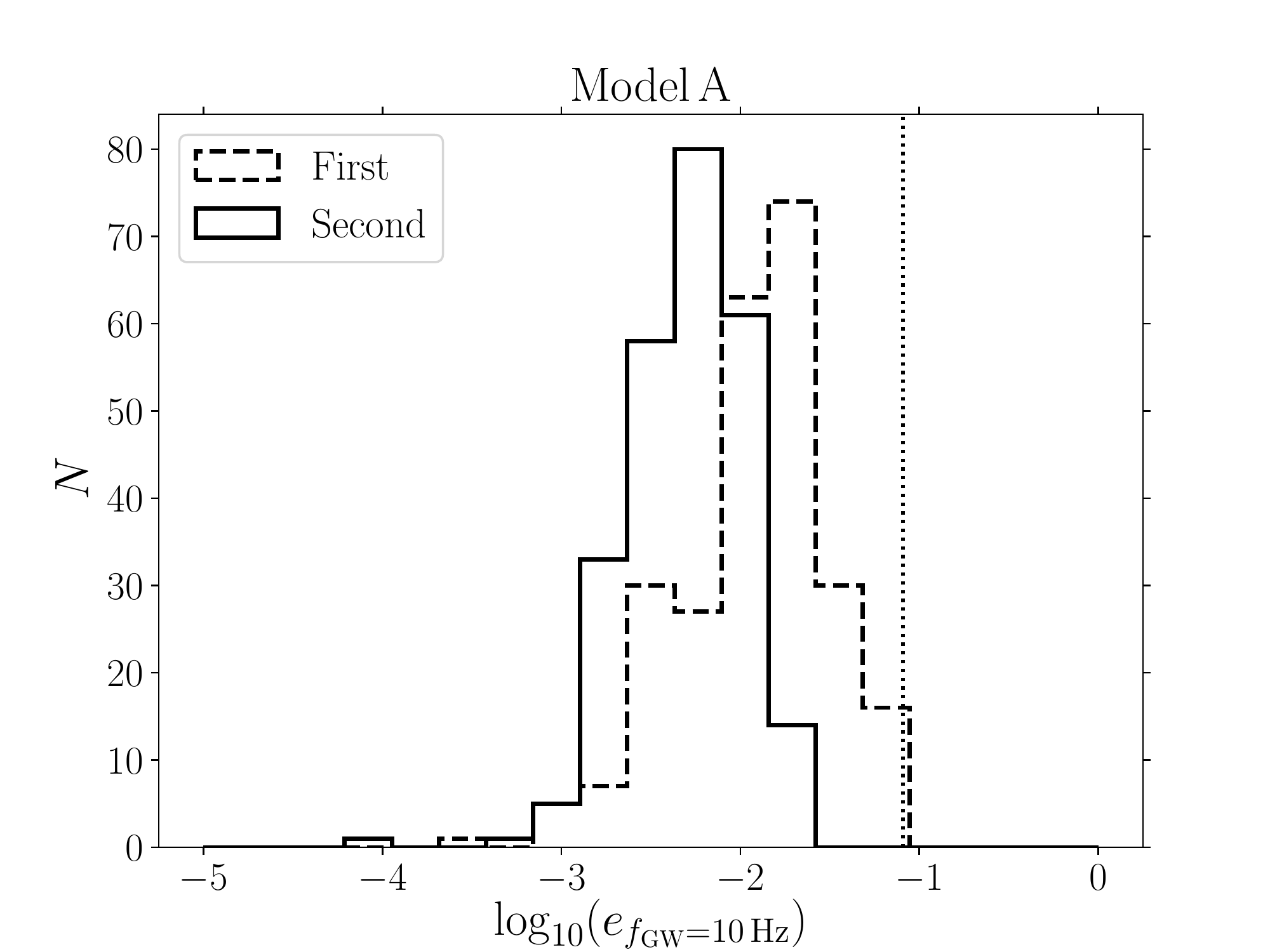}
}
\caption{ Distributions of the orbital eccentricities of the merging orbits (first and second mergers shown with dashed and solid lines, respectively) when the peak GW frequency is $f_\mathrm{GW}=f_\mathrm{GW,LIGO}=10\,\hz$. The black dotted vertical line shows the eccentricity above which LIGO/Virgo might distinguish between eccentric and circular sources \citep{2019ApJ...871..178G}. }
\label{fig:eL}
\end{figure}

After decoupling, these high eccentricities will dissipate due to GW emission. Nevertheless, some eccentricity remains when the orbital GW emission reaches the LIGO band ($f_\mathrm{GW,LIGO}=10\,\hz$). In \F~~\ref{fig:eL}, we show the distributions of the orbital eccentricities of the merging orbits (first and second mergers shown with dashed and solid lines, respectively) when the peak GW frequency is $f_\mathrm{GW}=f_\mathrm{GW,LIGO}$. Here, we evolve the orbit after decoupling using the equations of \citet{1964PhRv..136.1224P}, and calculate $f_\mathrm{GW}$ using \eq~(37) of \citet{2003ApJ...598..419W}. The remaining eccentricity when entering the LIGO band ranges between $\sim 10^{-3}$ and $\sim 10^{-1}$, and is slightly higher for the second mergers. In none of our systems, the eccentricity in the LIGO band exceeds $0.1$, which is approximately the eccentricity above which LIGO/Virgo might distinguish between eccentric and circular sources \citep{2019ApJ...871..178G}.

\subsubsection{Companion masses}
\label{sec:result:distr:mass}

\begin{figure}
\iftoggle{ApJFigs}{
\includegraphics[width=\linewidth]{test02_model_0_N_MC_100000_mode_chi_i_0_companion_masses}
}{
\includegraphics[width=\linewidth]{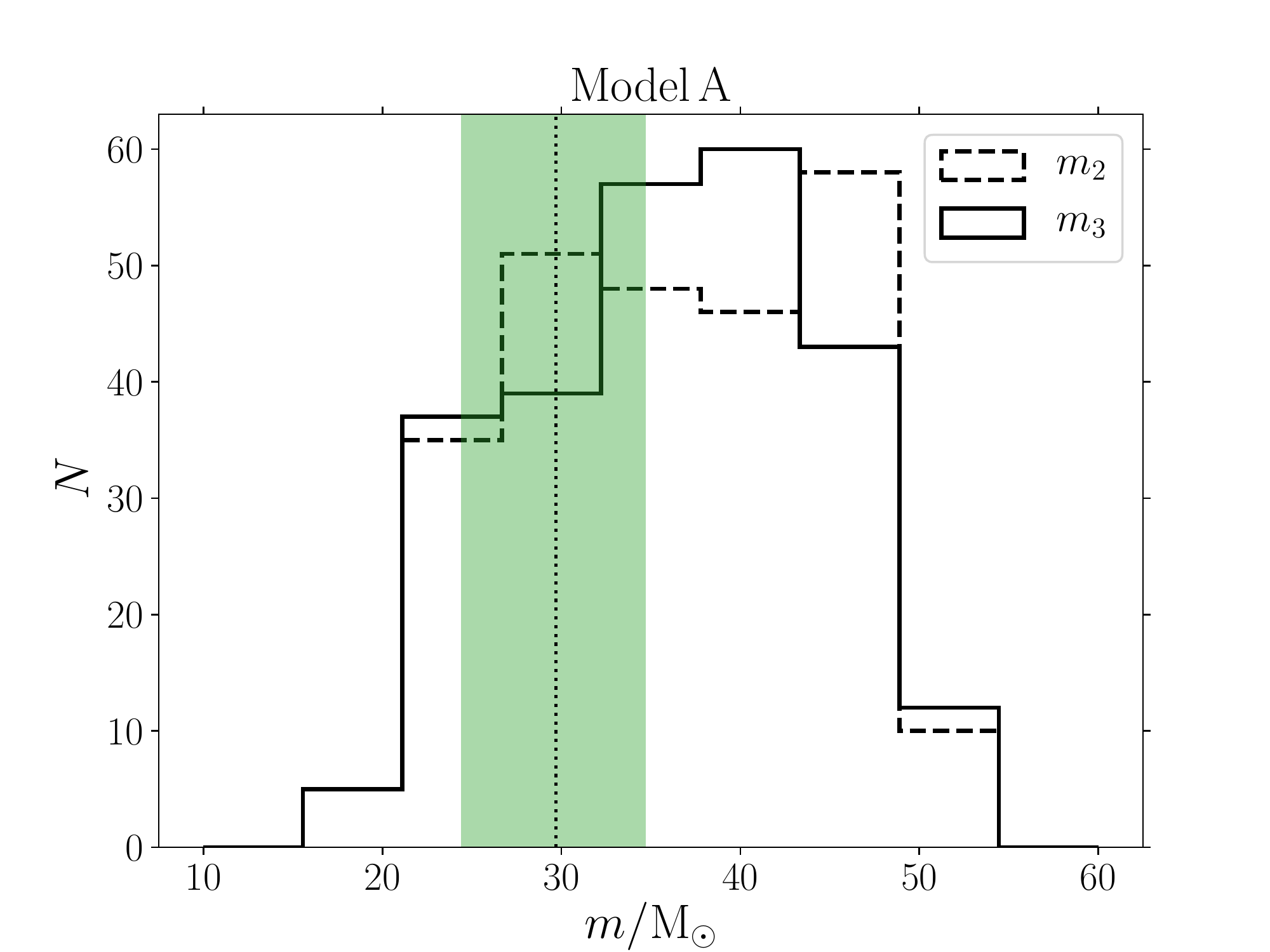}
}
\caption{ Distributions of $m_2$ (dashed line) and $m_3$ (solid line) for the double merger systems (model A). The primary mass of GW190412, $29.7_{-5.3}^{+5.0}\,\msun$ \citep{Abbottetal:2020tl}, is indicated with the black dotted line and green shaded region.  }
\label{fig:mass}
\end{figure}

In our simulations, the innermost two bodies have fixed masses, whereas $m_2$ and $m_3$ are sampled between $20$ and $50\,\msun$ (see \S\ref{sec:result:distr}). In \F~\ref{fig:mass}, we show the distributions of $m_2$ and $m_3$ for the double merger systems (model A). Both $m_2$ and $m_3$ for double merger systems are broadly distributed between the initial ranges, not showing any clear preferences. The primary mass of GW190412, $29.7_{-5.3}^{+5.0}\,\msun$ \citep{Abbottetal:2020tl}, is indicated in the figure, and is consistent with the distributions of $m_2$ in our simulations.

\subsubsection{Merger times}
\label{sec:result:distr:time}

\begin{figure}
\iftoggle{ApJFigs}{
\includegraphics[width=\linewidth]{test02_model_0_N_MC_100000_mode_chi_i_0_merger_times}
}{
\includegraphics[width=\linewidth]{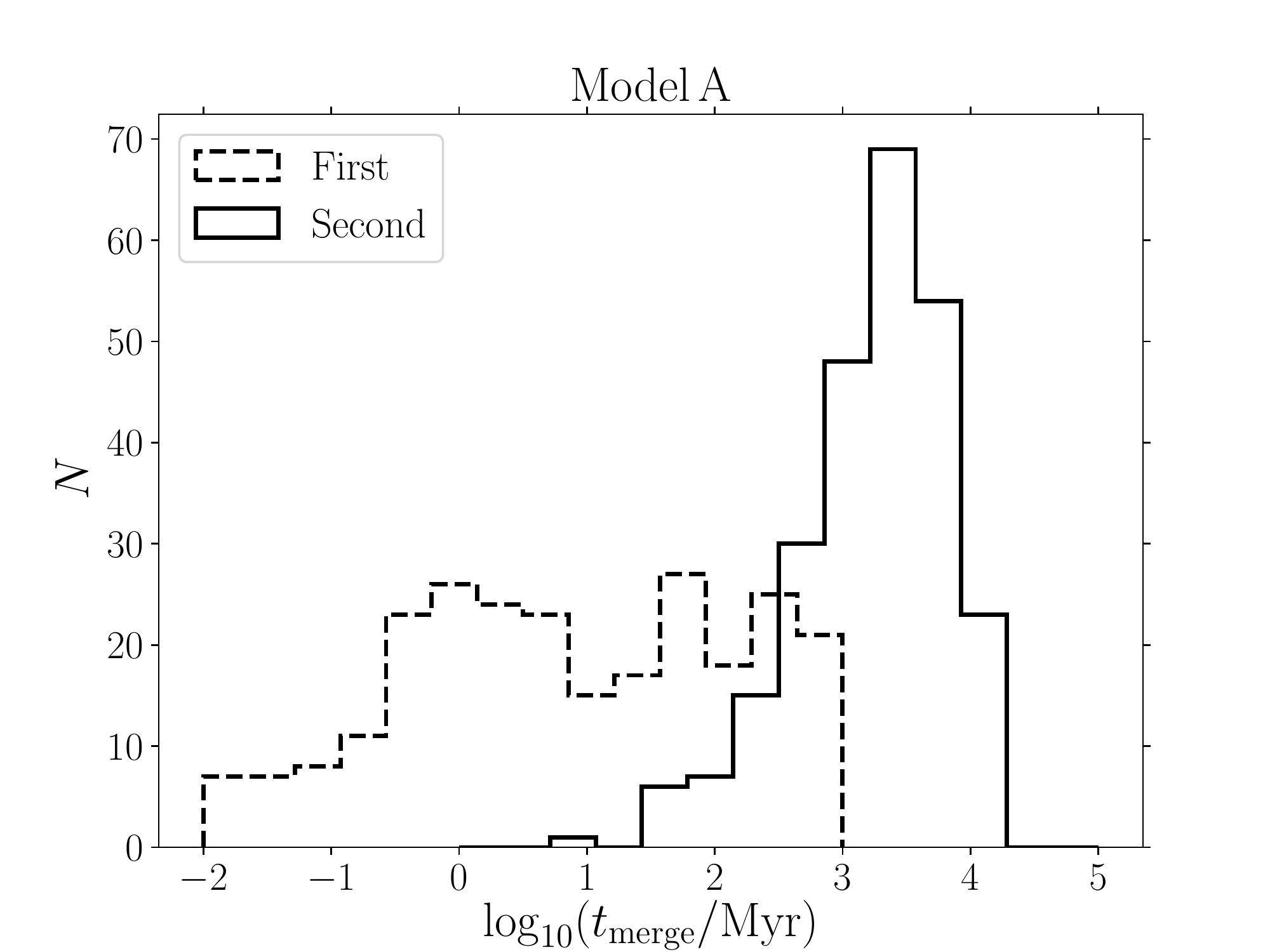}
}
\caption{ Distributions of the times of the first (dashed line) and second merger (solid line) for the double merger systems in our simulations (model A). }
\label{fig:time}
\end{figure}

Lastly, we show in \F~\ref{fig:time} the distributions of the times of the first (dashed line) and second merger (solid line) for the double merger systems in our simulations. We remind the reader that the simulations were run for up to $1\,\gyr$, and, if a first merged occurred, for an additional $14\,\gyr$ (cf. \S\ref{sec:model:meth}). 

The first merger occurs between a wide range of times between $\sim 10^{-2}$ and $10^3\,\myr$, and with a local peak around $1\,\myr$. This broad range is characteristic of the chaotic nature of the quadruple systems. The second merger typically occurs at significantly later times, and peaks around $\sim 3 \,\gyr$. This shows that double mergers can occur with significant delay times.

\section{Discussion}
\label{sec:discussion}
\subsection{Merger rates}
\label{sec:discussion:rate}
As discussed in \S\ref{sec:model:scen}, we ignored complications of quadruple evolution before the formation of four compact objects. Such complications include orbital expansion due to stellar evolution, CE evolution, and natal kicks. In order to obtain reliable double merger rates, a self-consistent study taking into account the stellar, binary, and dynamical evolution of 3+1 quadruple systems would need to be carried out, and this is left for future work. 

Nevertheless, we here briefly give a very rough back-on-the-envelope estimate of the double merger rate in 3+1 quadruples. We emphasize that this estimate is extremely uncertain, but include it nonetheless in order to make it plausible that at least some of the observed GW events might originate from double mergers in these systems.

We assume a local star formation rate per unit volume of $R_\star=10^{8}\msun\,\pgpy$ \citep{2014ARA&A..52..415M}. Next, based on the `canonical' model of \citet{2010MNRAS.402..519L} which is based on a Kroupa initial mass function \citep{2001MNRAS.322..231K}, we assume that the BH formation efficiency per unit Solar mass is $f_\mathrm{BH} = 0.01$, which leads to formation rate of $R_\mathrm{BH} = 10^6\,\pgpy$ BHs. The quadruple fraction among massive stars is significantly higher than for lower-mass stars \citep{2017ApJS..230...15M}; we adopt an optimistic fraction of $f_\mathrm{quad}=0.5$. In our integrations, we considered a subset of systems with specific orbital configurations. Most importantly, natal kicks, which are highly uncertain, could significantly reduce the number of `usable' systems. We assume that a fraction of $f_\mathrm{orb}=10^{-3}$ systems result in the orbital configurations adopted in our simulations. Lastly, the double merger fraction in our simulations is $\fdm \sim 10^{-3}$ (we ignore the complication that we use the local star formation rate, whereas double mergers can occur with significant delay times in the simulations). This gives a double merger rate of
\begin{align}
    R_\mathrm{dm} \sim f_\mathrm{quad} f _\mathrm{orb} \fdm R_\mathrm{BH} = 0.5\,\pgpy.
\end{align}
For reference, the LIGO/Virgo O1/O2 rate for BBH mergers is $53.2_{-28.2}^{+55.8}\,\pgpy$ \citep{2019ApJ...882L..24A}. Therefore, double mergers might account for a small fraction (a few per cent) of BBH mergers, although we stress again that the rate estimate given here is extremely uncertain. 

For comparison we mention a number of predictions of BBH merger rates from the literature. Predictions of the BH merger rate for triples include $\sim 1-30\,\pgpy$ \citep{2017ApJ...836...39S,2017ApJ...841...77A,2018ApJ...863....7R,2020MNRAS.493.3920F}, and, for quadruples, $\sim0.1\,\pgpy$ \citep{2019MNRAS.486.4781F}. Predictions for globular clusters include 2-20 $\mathrm{Gpc^{-3}\,yr^{-1}}$ \citep{2016PhRvD..93h4029R}, $>6.5 \,\mathrm{Gpc^{-3}\,yr^{-1}}$ \citep{2017MNRAS.469.4665P}, $>5.4 \,\mathrm{Gpc^{-3}\,yr^{-1}}$ \citep{2017MNRAS.464L..36A}, and 15-100 $\mathrm{Gpc^{-3}\,yr^{-1}}$ \citep{2018PhRvL.121p1103F}. For nuclear star clusters without massive BHs, the rates have been predicted to be $\sim 1 \, \mathrm{Gpc^{-3}\,yr^{-1}}$ \citep{2016ApJ...831..187A}; with massive BHs, rates estimates include $\sim0.1\,\gpc^{-3}\,\yr^{-1}$ \citep{2018ApJ...865....2H}, 1-3 $\,\gpc^{-3}\,\yr^{-1}$ \citep{2018ApJ...856..140H}, and $15\,\gpc^{-3}\,\yr^{-1}$ \citep{2017ApJ...846..146P}. 

\subsection{Future directions}
\label{sec:discussion:fut}
A closely related scenario for double mergers potentially producing GW190412-like sources to the one studied here could involve swapping the roles of bodies 0, 1 and 2: instead of assuming that $m_0,\,m_1\ll m_2$, assume that $m_0, \,m_1 \gg m_2$. The first merger would then consist of two BHs merging into a more massive BH, and the second merger would involve two highly unequal-mass BHs. The more massive BH in the second merger would likely have significant spin since it is a merger remnant, and this would help to explain the spin parameter of $\chi=0.43^{+0.16}_{-0.26}$ constrained for the primary (most massive) BH in GW190412 \citep{Abbottetal:2020tl}. This scenario, although potentially interesting, is beyond the scope of this work. 

In addition, as discussed in \S\ref{sec:model:scen}, pre-compact object evolutionary processes affect the formation of compact object 3+1 systems, and we ignored these complications here. Future work should include these processes in order to make more accurate predictions for the formation likelihood of GW190412-like systems through double mergers in 3+1 quadruples.

\section{Conclusions}
\label{sec:conclusions}

Three facts about GW190412 \citep{Abbottetal:2020tl} make it a unique BBH system: i) it has a low mass ratio ($q=0.28_{-0.07}^{+0.13}$), ii) it is a rather massive BBH system with a high effective spin ($\chie=0.25^{+0.09}_{-0.11}$) and a low false alarm rate, and iii) it is observed to have an in plane effective spin of $\chip =0.30^{+0.19}_{-0.15}$, which leads to precession of the orbit. 

The two main formation channels of BBH formation, isolated binary evolution and dynamical assembly in dense stellar systems, fail to explain GW190412. Here, we considered a scenario in which two mergers of compact objects occur in a wide hierarchical 3+1 quadruple system. First, two compact objects in the innermost orbit merge due to secular chaotic evolution. At a later time, the merged compact object coalesces with another compact object due to secular Lidov-Kozai oscillations. Our main conclusions are listed below.

\medskip \noindent 1. Based on population synthesis simulations of the dynamical evolution of 3+1 quadruples, we found that our scenario gives rise to distributions of the effective spin parameter $\chie$ and the in-plane spin parameter $\chip$ that are consistent with GW190412 (see \Fs~\ref{fig:chi_eff} and \ref{fig:chi_p}). The distributions of $\chie$ in our simulations are centered and symmetric around 0, and are driven by the changing orientation of the inner orbit of the triple after the first merger. The widths of the spin parameter distributions are mostly determined by the (uncertain) spin $\chi_2$ of the more massive component (see \S\ref{sec:result:distr:spin}, and Appendix~\ref{app:spin}). We find that spin-orbit terms after the first merger are not important. The distribution of $\chip$ in our scenario is typically broadly distributed but with some preference for $\chip$ around 0.2, consistent with GW190412.

\medskip \noindent 2. Assuming that GW190412 originated from a double merger in a compact object 3+1 quadruple system, we find a strong constraint that the first merger likely occurred between approximately equal-mass BHs in the innermost orbit, since the recoil velocity from unequal-mass BHs would otherwise have disrupted the system. This is based on the fact that, although rare, double mergers can occur when the innermost masses are equal ($m_0=m_1=4\,\msun$; model D), whereas no mergers occurred in our simulations when the recoil velocity was taken into account with systems with innermost masses of $m_0=6.5\,\msun$ and $m_1=1.5\,\msun$ (model E; see Table~\ref{table:models}). 

\medskip \noindent 3. In our scenario, extremely high eccentricities need to be reached in order to achieve eccentricity-boosted mergers in relatively wide orbits (see \S\ref{sec:result:distr:orbit}). However, the eccentricities when reaching the LIGO detector band are not high enough to be currently distinguishable. This may change in the future, when next-generation GW detectors become more sensitive, and the theoretical modelling of eccentric GW waveforms improves.

\medskip \noindent 4. Our model has a wide range of delay times for the second merger, with long delay times of up to a Hubble time being possible (see \S\ref{sec:result:distr:time}). Therefore, systems merging at low redshift could have formed at much higher redshift. 

\acknowledgements 
We thank the anonymous referee for a helpful report. This work is supported by the National Science Foundation under Grant No. AST-1440254.

\bibliographystyle{apj}
\bibliography{literature.bib}

\appendix


\section{Spin parameter distributions for different assumptions on $\chi_2$}
\begin{figure*}
\iftoggle{ApJFigs}{
\includegraphics[width=0.49\linewidth]{test02_N_MC_100000_mode_chi_i_1_chi_eff}
\includegraphics[width=0.49\linewidth]{test02_N_MC_100000_mode_chi_i_2_chi_eff}
}{
\includegraphics[width=0.49\linewidth]{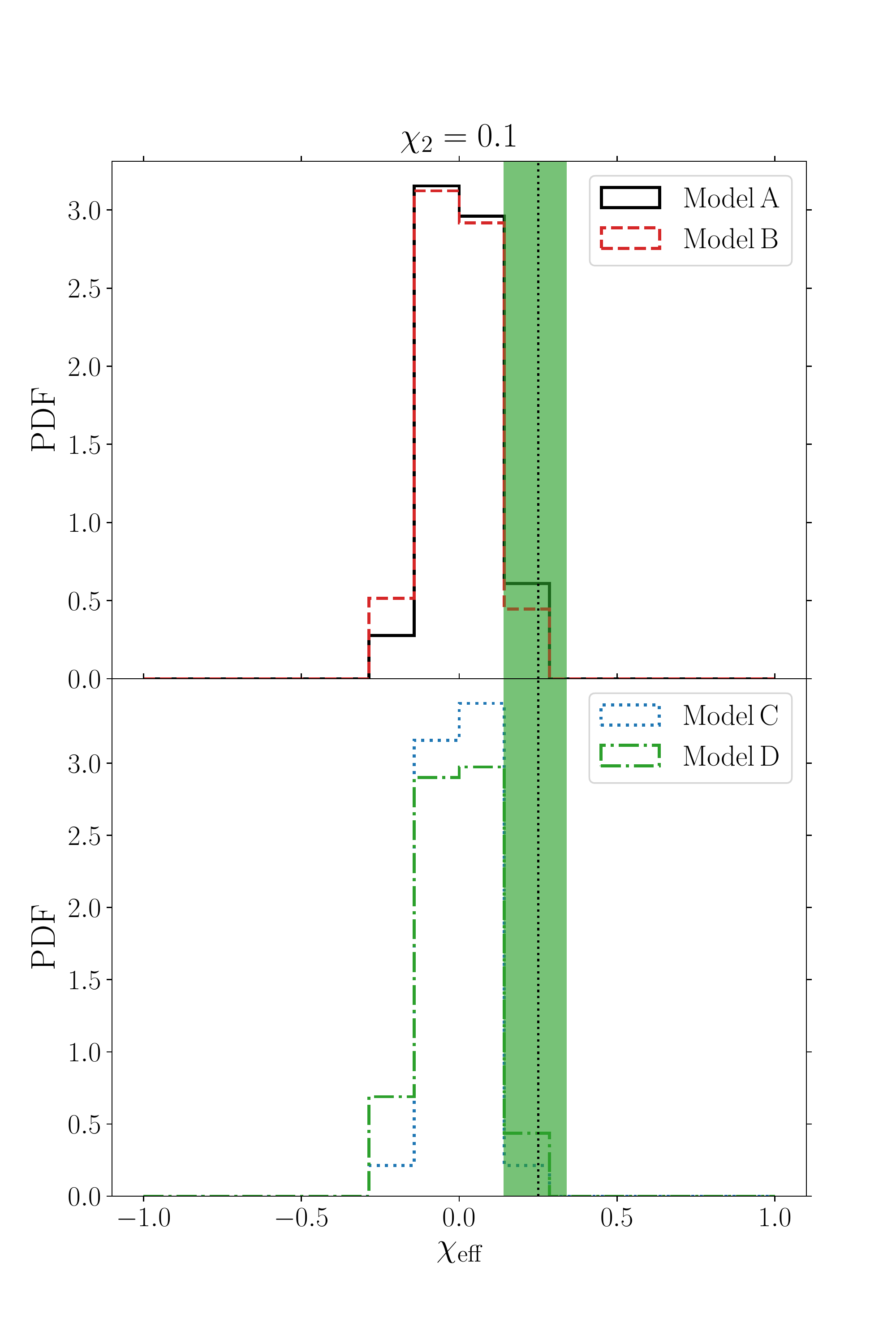}
\includegraphics[width=0.49\linewidth]{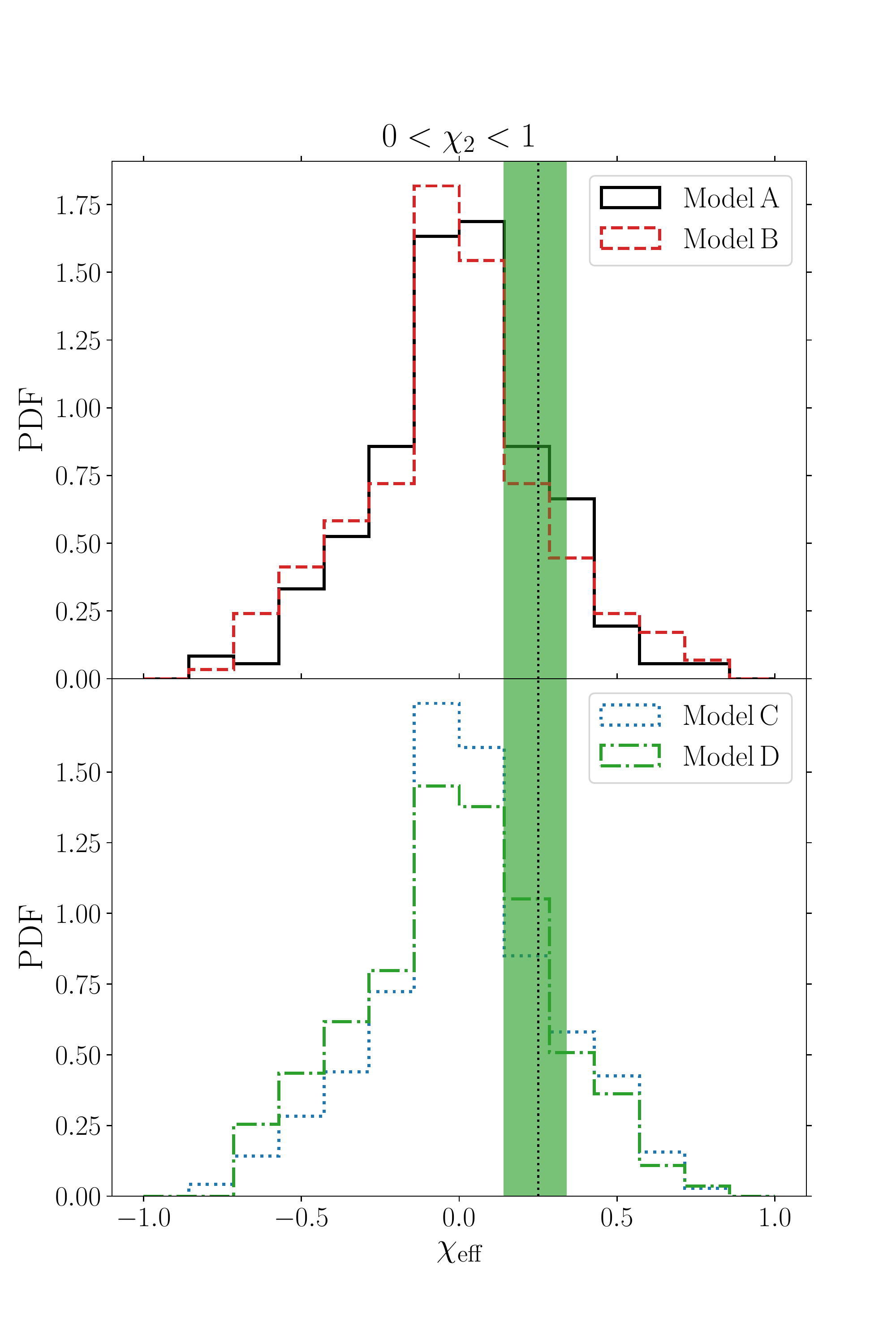}
}
\caption{ Distributions of $\chie$ for the double mergers in our simulations, similar to \F~\ref{fig:chi_eff}, but here setting either $\chi_2=0.1$ (left-hand panels), or $\chi_2$ sampled uniformly between 0 and 1 (right-hand panels). The LIGO/Virgo value for GW190412, $\chie=0.25_{-0.11}^{+0.09}$ \citep{Abbottetal:2020tl}, is shown with the dotted black lines and green shaded regions. }
\label{fig:chi_eff_app}
\end{figure*}

\begin{figure*}
\iftoggle{ApJFigs}{
\includegraphics[width=0.49\linewidth]{test02_N_MC_100000_mode_chi_i_1_chi_p}
\includegraphics[width=0.49\linewidth]{test02_N_MC_100000_mode_chi_i_2_chi_p}
}{
\includegraphics[width=0.49\linewidth]{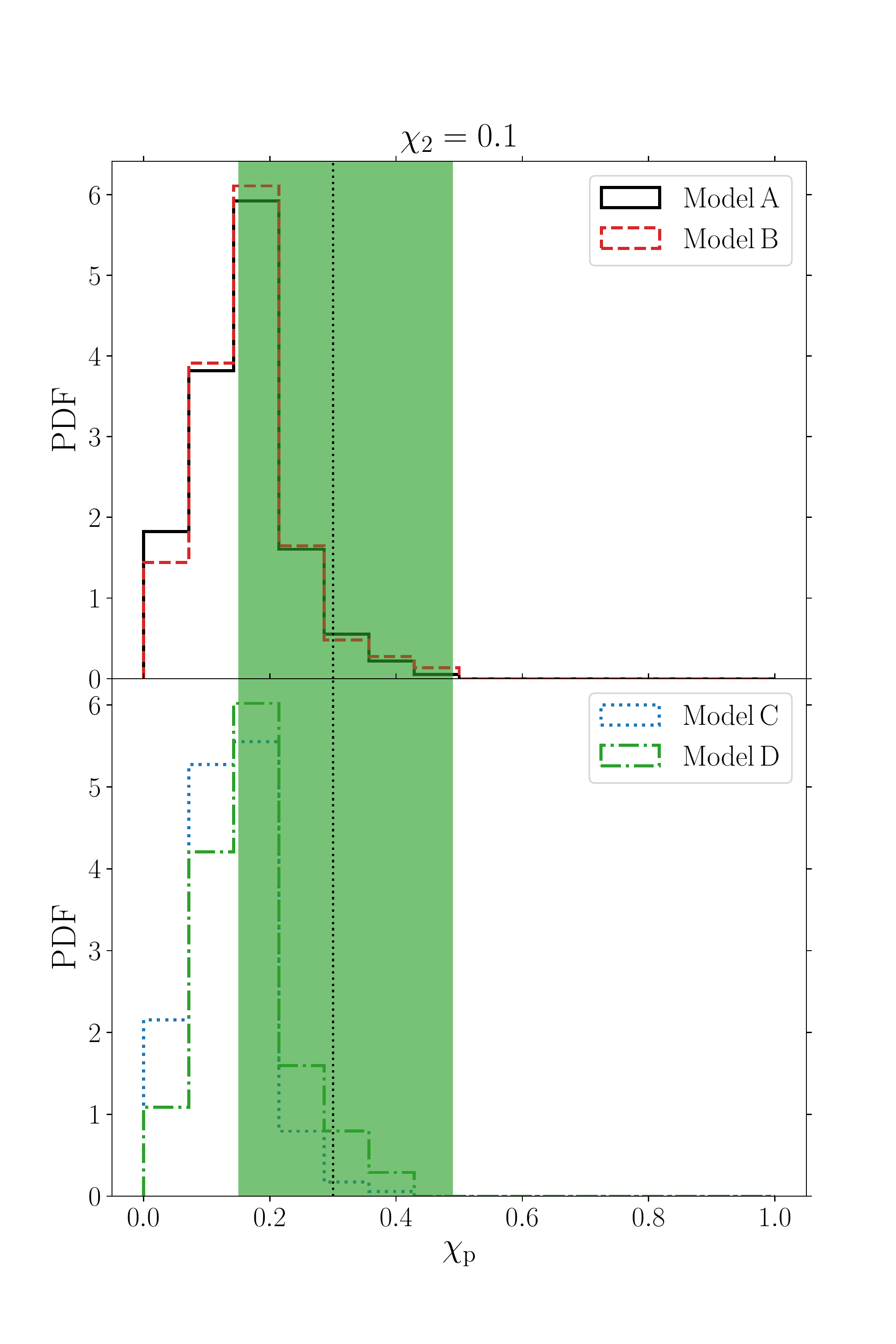}
\includegraphics[width=0.49\linewidth]{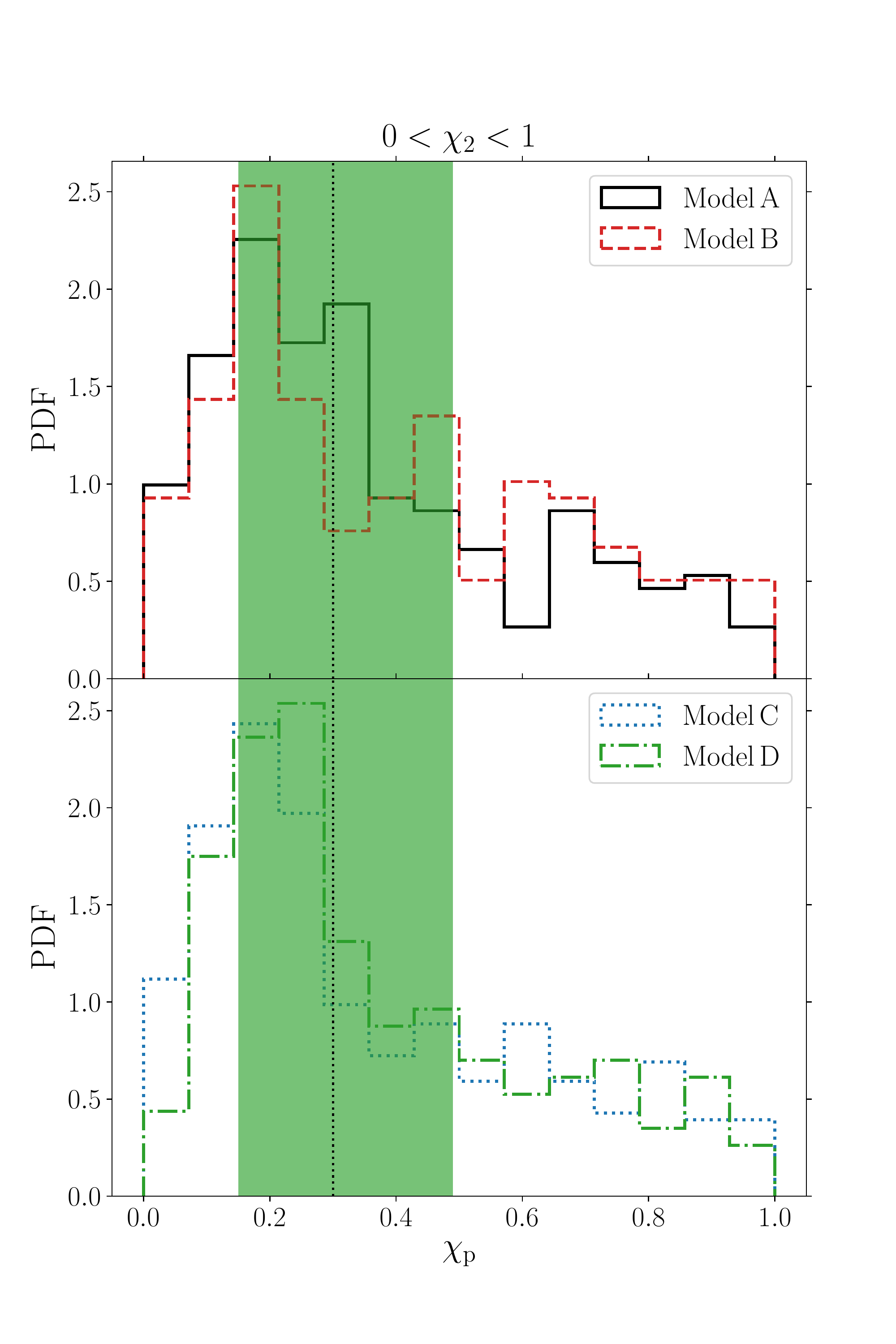}
}
\caption{ Distributions of $\chip$ for the double mergers in our simulations, similar to \F~\ref{fig:chi_p}, but here setting either $\chi_2=0.1$ (left-hand panels), or $\chi_2$ sampled uniformly between 0 and 1 (right-hand panels). The top (bottom) panels show distributions for models A and B (C and D), with line styles and colors as indicated in the legend. The LIGO/Virgo value for GW190412, $\chip=0.30_{-0.15}^{+0.19}$ \citep{Abbottetal:2020tl}, is shown with the dotted black lines and green shaded regions. }
\label{fig:chi_p_app}
\end{figure*}

In \Fs~\ref{fig:chi_eff_app} and \ref{fig:chi_p_app}, we show the distributions of $\chie$ and $\chip$, respectively, with different assumptions on the value of $\chi_2$. 

\label{app:spin}

\end{document}